# Moderating model marketplaces: platform governance puzzles for AI intermediaries

Robert Gorwa[a] and Michael Veale[b,c]

[a]WZB Berlin Social Science Center, Berlin, Germany; [b]Faculty of Laws, University College London, London, UK; [c]Institute for Information Law, University of Amsterdam, Amsterdam, Netherlands

**ABSTRACT**
The AI development community is increasingly making use of hosting intermediaries, such as Hugging Face, which provide easy access to user-uploaded models and training data. These *model marketplaces* lower technical deployment barriers for hundreds of thousands of users, yet can be used in numerous potentially harmful and illegal ways. In this article, we explain the ways in which AI systems, which can both 'contain' content and be open-ended tools, present one of the trickiest platform governance challenges seen to date. We provide case studies of several incidents across three illustrative platforms – Hugging Face, GitHub and Civitai – to examine how model marketplaces moderate models. Building on this analysis, we outline important (and yet nevertheless limited) practices that industry has been developing to respond to moderation demands: licensing, access and use restrictions, automated content moderation, and open policy development. While the policy challenge at hand is a considerable one, we conclude with some ideas as to how platforms could better mobilise resources to act as a careful, fair, and proportionate regulatory access point.

## 1. Introduction

In the summer of 2022, a Swiss machine learning researcher uploaded a video subtitled '*This is the Worst AI Ever*.' In a nineteen-minute clip that quickly began to amass tens of thousands of views, the YouTuber Yannic Kilcher describes how he created a new model, jokingly called

'GPT-4chan,' by combining an existing open-source large language model (EleutherAI's GPT-3–like 'GPT-J') with a published dataset of over 130M posts scraped from 4chan's 'politically incorrect' /pol/ imageboard.[1] With apparent pleasure, Kilcher described a 'prank' he had run, using the model to power a chatbot that he then let loose on /pol/. This large-language model for hate speech would post anonymously more than 30,000 times on 4chan before eventually being shut off by its creator.[2]

Alongside his video, Kilcher uploaded GPT-4chan to Hugging Face, a rapidly growing platform for open-source AI development. Via Hugging Face, visitors arriving from Kilcher's YouTube page or elsewhere could, with a single click, download the model along with instructions for its deployment, or interact with it through the platform's built in 'playground' cloud environment.

Tuning a powerful generative text model on interactions from the world's most infamous online 'cultural breeding ground for far-right hate and violent extremism'[3] yielded unsurprisingly toxic results. As one graduate student posting on the Hugging Face community pages noted:

> I tried out the demo mode of your tool 4 times, using benign tweets from my feed as the seed text. In the first trial, one of the responding posts was a single word, the N word. The seed for my third trial was, I think, a single sentence about climate change. Your tool responded by expanding it into a conspiracy theory about the Rothchilds and Jews being behind it.[4]

Kilcher's stunt quickly stirred up controversy in the American and European machine learning communities. A number of high-profile researchers argued 'that the model has already or is very likely to cause harm, that making the bot interact with 4chan users was unethical, and that Kilcher knew this would cause controversy and did all this with the specific intent for that to happen.'[5]

It also kicked Hugging Face's small staff into gear, as they began to discuss the model on the site's community 'talk' pages – a mini forum attached to each model repository. Three days after GPT-4chan went online, CEO Clément Delangue intervened directly, noting that he did not 'support the training and experiments done […] with this model,' which were 'IMO pretty bad and inappropriate,' but that Hugging Face was working on its 'ethical review' processes and could allow the model to remain online if Kilcher provided more disclaimers about its issues/limitations and raised the barrier-to-entry for less technical members of the public by disabling the model's interactive 'playground' and one-click deployment features.[6] The platform did not yet have a content policy which would come into play in a scenario like this, let alone a structured 'trust and safety' bureaucracy like most conventional user-generated content intermediaries.[7] Working through these issues, and despite Delangue's initial comments, Hugging Face staff eventually decided to block access to GPT-4chan completely.

---

[1] Antonis Papasavva and others, 'Raiders of the Lost Kek: 3.5 Years of Augmented 4chan Posts from the Politically Incorrect Board' in *Proceedings of the Fourteenth International AAAI Conference on Web and Social Media (ICWSM 2020)* (2020) <https://doi.org/10.1609/icwsm.v14i1.7354>.
[2] James Vincent, 'YouTuber trains AI bot on 4chan's pile o' bile with entirely predictable results' (*The Verge*, June 2022) <www.theverge.com/2022/6/8/23159465/youtuber-ai-bot-pol-gpt-4chan-yannic-kilcher-ethics> accessed 16 July 2023.
[3] Sal Hagen, '"Who is/Ourguy/?": Tracing Panoramic Memes to Study the Collectivity of 4chan/Pol' (2022) *New Media & Society* 1735. <https://doi.org/10.1177/14614448221078274>, 2.
[4] <https://huggingface.co/ykilcher/gpt-4chan/discussions/1#629ebdf246b4826be2d4c8c9> (archived at <https://perma.cc/JDZ8-JR4C>).
[5] Andrey Kurenkov, 'Lessons from the GPT-4Chan controversy' (*The Gradient*, June 2022) <https://thegradient.pub/gpt-4chan-lessons/> accessed 16 July 2023.

Not all users were satisfied with this decision. One anonymous user noted disapprovingly that 'a model is a tool,' insinuating that generative AI systems inherently had ambivalent and dual-use valences.[8] How could a platform properly police how others decided to use open-source tools – tools which could be feasibly used for everything from legitimate research into toxic-speech detection to targeted hate and harassment campaigns? And how *should* a company do so from both a legal and ethical standpoint?

## 1.1. The platformisation of AI (and its governance)

The global AI ecosystem is becoming platformised across multiple dimensions.[9] Yet in contrast to early predictions that powerful machine learning tools would only be deployed by wealthy actors with requisite technical sophistication and access to training data and computational power, a notable trend has been the public release of leading edge models under various 'open-source' licenses.[10] Crucial to this development has been the emergence of AI development intermediaries that we call 'model marketplaces,' epitomised by the New York-based start-up Hugging Face – as well as other competing platforms offering related functionality, such as Replicate and GravityAI. The earliest of these firms were founded in the

---

[6]<https://huggingface.co/ykilcher/gpt-4chan/discussions/1#629e6d4abb6419817edfb1d7> (archived at <https://perma.cc/JDZ8-JR4C>).
[7]Kate Klonick, 'The New Governors: The People, Rules, and Processes Governing Online Speech' (2017) 131(6) *Harvard Law Review* 1598; Sarah T Roberts, *Behind the Screen: Content Moderation in the Shadows of Social Media* (Yale University Press 2019).
[8]<https://huggingface.co/ykilcher/gpt-4chan/discussions/4#62a2ca7103bf94c3ac52707c> (archived at <https://perma.cc/CAH6-BUZ6>).
[9]Jennifer Cobbe and Jatinder Singh, 'Artificial Intelligence as a Service: Legal Responsibilities, Liabilities, and Policy Challenges' (2021) 42 *Computer Law & Security Review* 105573 <https://doi.org/10/gmq8jm>.
[10]However, many of these models are neither using true open-source licenses, nor does their apparent openness dilute power centralisation and conslidation in this ecosystem. See generally David Gray Widder, Sarah West, and Meredith Whittaker, 'Open (For Business): Big Tech, Concentrated Power, and the Political Economy of Open AI' (SSRN, August 2023) <https://doi.org/10.2139/ssrn.4543807> accessed 29 August 2023.

mid-2010s as services where developers could upload AI systems and receive royalties when they were accessed, traded or used.[11]

Model marketplaces are a new form of user-generated content platform, where users can upload AI systems and AI-related datasets, which in turn can be downloaded, and depending on the business model, queried, tweaked, or built upon by other users. They are related to generic software development platforms like GitHub or GitLab – so much so that we analyse GitHub in this paper as a model marketplace – but are notable for developing new features beyond classic software repositories, such as model querying, deployment, and user interfaces.

As with any other user-generated content platform, therefore, there are many conceivable ways in which model marketplaces can be – and already are – being used for nefarious ends. From 'pranks' involving the development of potentially dangerous large-language infrastructures for hate and harassment like GPT-4chan to models that create synthetic yet realistic non-consensual pornography, these open-source AI platforms are now facing a whole spectrum of old and new 'trust and safety' issues. In this paper, we explore how the companies that operate model marketplaces are slowly developing policies regarding models that have an explicitly political valence, that engage in satire, that defame people, that can create text and images depicting illegal or otherwise socially problematic behaviour (child abuse imagery, terrorist content), that infringe copyright, and more.[12] These companies are also beginning to interface with established policy frameworks and experiencing pressure from governance stakeholders seeking to obtain the removal of certain models. In other words, model marketplaces are now grappling with the kinds of difficult questions that have in recent years been explored in a large interdisciplinary literature on 'platform governance' and content moderation in the social media context and beyond.[13]

However, models are also inherently more complex to govern than traditional forms of user-generated material. Models can be used to do things in the world that traditional content does not or cannot. They have complex affordances, which can be understood following a socio-technical research tradition as akin to those of ambivalent dual-use technologies,[14] with the same

---

[11]Davey Alba, 'Need some AI? Yeah, there's a marketplace for that' (*Wired*, September 2016) <https://www.wired.com/2016/09/algorithmia-deep-learning/> accessed 28 August 2023.

[12]Susan Hao and others, 'Safety and Fairness for Content Moderation in Generative Models' (arXiv June 2023) <http://arxiv.org/abs/2306.06135> accessed 29 August 2023.

[13]Tarleton Gillespie, *Custodians of the Internet: Platforms, Content Moderation, and the Hidden Decisions That Shape Social Media* (Yale University Press 2018); Robert Gorwa, 'What is Platform Governance?' (2019) 22(6) *Information, Communication & Society* 854 <https://doi.org/10.1080/1369118X.2019.1573914>; Robyn Caplan, 'Networked Platform Governance: The Construction of the Democratic Platform' (2023) 17 *International Journal of Communication* 3451.

[14]Whitney Phillips and Ryan M Milner, *The Ambivalent Internet: Mischief, Oddity, and Antagonism Online* (Polity, 2018); Jonathan B Tucker, *Innovation, Dual Use, and Security: Managing the Risks of Emerging Biological and Chemical Technologies* (MIT Press, 2012); Peter Henderson and others, 'Self-Destructing

core technical infrastructure permitting not just civilian/research/benign applications but also far more damaging/surveillant/military applications. Although models can be uploaded by third parties, and hosted and made accessible via platforms like GitHub or Hugging Face that then govern the visibility, ease of access, and other dimensions of how others interface with these models, considerations of governing AI models as content need to move beyond discussions that only seek to balance trade-offs between free expression and public safety, as in the classic content moderation context.

Models are tools. In certain cases, they can be particularly powerful tools that can be used in the real world in ways more damaging than single instances of 'harmful content.' For generative models, this includes, but is not limited to, the especially rapid and low-cost production of things like non-consensual pornography, disinformation, or even incitement to violence.[15] The stakes are high: while a single image or video may cause distress to the individuals that see it, a generative model could feasibly be plugged into automated systems that spread its output (which could be anything from spam to computer-generated child abuse imagery) across the internet. Complicating matters further, these models need not be explicitly tuned to for harm from the outset: savvy actors could deploy such systems by building upon freely available general-purpose technologies. How should the new platforms in this space consider the prospective downstream effects of the models that they provide public access to (and, depending on their business model, profit from)? How could and are these companies grappling with the emerging governance challenges that they face?

We begin with a historically informed overview of the evolution and rise to prominence of model marketplaces. Section 2 offers a brief exploration of key actors, features, and business models that have become an integral part of the AI development/deployment intermediary ecosystem. We then turn to a conceptual survey of the particularly challenging features that models on these marketplaces exhibit, which – we argue – motivate various policy questions. Firstly, models contain content – and generative models, in particular, have a tendency to memorise and/or semantically reconstruct content which they have ingested during training. This potentially exposes developers and model marketplace operators to both formal legal liability and informal pressure from policy stakeholders (such as copyright holders or government security actors). Secondly, models are tools which can be used in a variety of unexpected, generative, dual-use, and potentially harmful ways by third-

---

Models: Increasing the Costs of Harmful Dual Uses of Foundation Models' in *Proceedings of the 2023 AAAI/ACM Conference on AI, Ethics, and Society* (AIES '23: AAAI/ACM Conference on AI, Ethics, and Society, ACM, 8 August 2023) <https://doi.org/10.1145/3600211.3604690>.

[15]Laura Weidinger and others, 'Taxonomy of Risks Posed by Language Models' in *Proceedings of the 2022 ACM Conference on Fairness, Accountability, and Transparency* (FAccT '22, ACM June 2022) <https://doi.org/10.1145/3531146.3533088>.

party actors downstream. Despite these challenges, model marketplaces currently exist in a sort of regulatory vacuum, and fall through the cracks both of the leading platform regulation and AI regulation frameworks. As of right now, these platforms are operating in a self-regulatory mode without clear legal or policy guidance.

Section 3 offers short case studies to look at how some exemplary open-source AI development intermediaries have dealt with these puzzles in recent years. Drawing on an analysis of platform policies, terms of service documents, community guidelines, and relevant discussion pages, forums, and other venues through which these platforms articulate their decision-making publicly, we look at the active efforts by Hugging Face to govern certain models as content, and examine how its effort to deal downstream with the potential impact of the models they provide access to has been tied to their promotion of licensing as a potential governance mechanism. We then examine GitHub (a more generic developer platform which nonetheless offers access to many machine learning models), discussing how it has sought to develop policies for software exploits and dual-use hacking tools, and how it has more recently sought to handle the proliferation of models for 'deepfake' synthetic content generation via its service. We also look at a less-mature but rapidly growing platform, the Civitai marketplace for image generation models, and explore how it has dealt with the plethora of non-consensual sexual imagery and copyright infringing content being created via the tools that users are offering on its platform.

The article closes with a discussion of the key emerging governance modalities in this ecosystem. In particular, we discuss how model marketplaces are developing new norms around model gating and access, incentivising actors to embed safety systems (many of which are quite flawed) into their value chain, and developing creative but potentially unsustainable moderation practices around AI model licensing.

## 2. Understanding model marketplaces

### 2.1. Actors, affordances, business models

Programmers, hackers, tinkerers, and all other sorts of computer hobbyists have always shared and exchanged software. Early digitally mediated examples of this practice include everything from UNIX customisation files being shared via email,[16] to programs and other media being posted to bulletin board systems.[17] It was not until the late 1990s, however, that these practices were platformised and institutionalised via services that hosted, structured, and facilitated community access to software. Source-Forge, launched by the Californian company VA Software in 1999, provided open-source projects with free codebase hosting, version control, and community communication and collaboration tools.[18] GitHub, which allows users to host code repositories, 'fork' and tweak the repositories of others, and streamline collaboration on projects via version control tools,[19] launched in 2008 and became enormously popular for source code maintenance and other aspects of software development before being acquired by Microsoft in 2018.

---

[16] Wendy E Mackay, 'Patterns of Sharing Customizable Software' in *Proceedings of the 1990 ACM Conference on Computer-Supported Cooperative Work* (CSCW '90, ACM, 1990).
[17] Ville Oksanen and Mikko Välimäki, 'Theory of Deterrence and Individual Behavior. Can Lawsuits Control File Sharing on the Internet?' (2007) 3(3) *Review of Law & Economics* 693 <https://doi.org/10.2202/1555-5879.1156>.

There are many different actors engaged in the 'applied science and engineering discipline'[20] that is commonly termed 'artificial intelligence.' As what we might now consider the ongoing 'AI summer' began heating up in the early and mid-2010s, a few start-ups sought to combine some of the technical features of these aforementioned open-source software development platforms with marketplace dynamics that brought together these different actors. A notable early player was Algorithmia, which was founded in 2013 in Seattle by former Microsoft engineers. The company advertised itself as the provider of an 'open marketplace for algorithms,' where third-party developers could upload 'working algorithms designed to slot right into new services.'[21] In 2018, coverage of Algorithmia in the business press boasted that the platform had 'over 60,000 developers tapping into a library of over 4,500 algorithms.'[22] In 2021, it was acquired by Boston-based enterprise AI firm DataRobot.[23]

Although Algorithmia did feature some open-source models, its business model was primarily a multi-sided one, with the company taking a transaction fee whenever the third-party models that they hosted were queried via the Algorithmia API.[24] Microsoft appears to have offered a similar service via its Cortana Intelligence Gallery (later renamed the Azure AI Gallery), again targeting enterprise customers looking for the easy integration of new systems, and the long-term maintenance of those tools once integrated in a corporate environment. In other words, these services were being developed for, and branded as part of, the emerging practice of 'MLOps' – machine learning operations – which grew out from the notion of 'DevOps' (or development operations) in software engineering. MLOps is a set of practices seeking to effectively implement machine learning systems inside complex manufacturing, production, IT, or other systems, and seeking to do so in a monitorable and reproducible manner: it involves the 'coordination of the resulting, often complex ML system components and infrastructure, including the roles required to automate and operate an ML system in a real-world setting.'[25]

---

[18]Damian Andrew Tamburri and others, '"The Canary in the Coal Mine..." A Cautionary Tale from the Decline of SourceForge' (2020) 50(10) *Software: Practice and Experience* 1930 <https://doi.org/10.1002/spe.2874>.
[19]Laura Dabbish and others, 'Social Coding in GitHub: Transparency and Collaboration in an Open Software Repository' in *Proceedings of the ACM 2012 Conference on Computer Supported Cooperative Work* (CSCW '12, ACM, 2012) <https://doi.org/10.1145/2145204.2145396>.
[20]Joanna J Bryson, 'The Artificial Intelligence of the Ethics of Artificial Intelligence: An Introductory Overview for Law and Regulation' in Markus D Dubber, Frank Pasquale, and Sunit Das (eds), *The Oxford Handbook of Ethics of AI* (Oxford University Press, 2020) <https://doi.org/10.1093/oxfordhb/9780190067397.013.1> 4.
[21]Alba (n 11).
[22]Amit Chowdhry, 'How Algorithmia Built the Largest Marketplace for Algorithms in the World' (*Forbes*, January 2018) <www.forbes.com/sites/amitchowdhry/2018/01/22/how-algorithmia-built-the-largest-marketplace-for-algorithms-in-the-world/> accessed 14 September 2023.
[23]Taylor Soper, 'DataRobot lands $300M and acquires Seattle machine learning startup algorithmia' (*GeekWire*, July 2021) <www.geekwire.com/2021/datarobot-lands-300m-acquires-seattle-machine-learning-startup-algorithmia/> accessed 14 September 2023.
[24]Alba (n 11).

Hugging Face – founded in New York in 2016 as a computational linguistics start-up – managed to break through in the early 2020s as the machine learning development platform most widely used both by researchers and in industry. After an initial foray into the development of natural language processing-fuelled chatbots, it pivoted towards creating a product that is commonly described as 'the GitHub of machine learning.'[26] Oriented initially towards the research community, Hugging Face created a free platform through which third-parties could access and/or share datasets, software libraries, and pretrained models.[27] It has also managed to fill a niche as part of a MLOps pipeline, integrating its repositories with ML-deployment infrastructures like Amazon SageMaker. This has been facilitated by its release of its open-source 'Transformers' library,[28] which can be used to achieve interoperability across leading development frameworks like PyTorch (released by Facebook AI Research) and TensorFlow (developed by Google).

In a book-length trade publication providing an in-depth look at the ways that organisations and individuals can use their platform, Hugging Face researchers describe what they see as the core added value of their 'model hub':

> In the early days, pretrained models were just posted anywhere, so it wasn't easy to find what you needed. Murphy's law guaranteed that PyTorch users would only find TensorFlow models, and vice versa. And when you did find a model, figuring out how to fine-tune it wasn't always easy. This is where

---

[25] Dominik Kreuzberger, Niklas Kühl, and Sebastian Hirschl, 'Machine Learning Operations (MLOps): Overview, Definition, and Architecture' (2023) 11 *IEEE Access* 31866 <https://doi.org/10.1109/ACCESS.2023.3262138>, 31866.

[26] Faustine Ngila, 'The GitHub of AI is named after an emoji — and Microsoft has its fingers in it already' (*Quartz*, May 2023) <https://qz.com/hugging-face-microsoft-artificial-intelligence-1850490270> accessed 27 September 2023.

[27] Emilia David, 'Google, Amazon, Nvidia, and Others put $235 million into hugging face' (*The Verge*, August 2023) <www.theverge.com/2023/8/24/23844444/google-amazon-nvidia-hugging-face-generative-ai-investment> accessed 25 August 2023.

[28] Thomas Wolf and others, 'HuggingFace's Transformers: State-of-the-art Natural Language Processing" (arXiv, July 2020) <https://doi.org/10.48550/arXiv.1910.03771> accessed 27 September 2023.

> Hugging Face's Transformers library comes in: it's open source, it supports both TensorFlow and PyTorch, and it makes it easy to download a state-of-the-art pretrained model from the Hugging Face Hub, configure it for your task, fine-tune it on your dataset, and evaluate it.[29]

Hugging Face has grown rapidly, hosting over 300,000 model repositories as of August 2023, with purportedly more than 4 million downloads of its Transformers library every month.[30] The most downloaded models as of August 2023 involve a mix of professional and non-professional content, with the top 10 models including a fine-tuned version of a Facebook speech recognition model uploaded by a Brazilian PhD student, a few large language models ranging from OpenAI's GPT-2 to Facebook's LLaMA, and version 1.5 of Stable Diffusion's image generation model (Figure 1). Major research organisations, university institutes, and industry labs all disseminate their work directly on Hugging Face, making the platform (a) an important path to content discovery and distribution in the machine learning space, and (b) a potentially influential 'chokepoint' or gatekeeper for future AI policy efforts.

Even as it has grown rapidly, the firm has sought to cultivate a reputation as a responsible actor in the 'fair and ethical' AI ecosystem. Hugging Face researchers have conducted some important critical research on bias in image generation models,[31] and played an important role orchestrating notable open-source science projects like Bloom.[32] The firm has also voluntarily integrated some basic yet nevertheless notable transparency features into its platform design (such as the 'model cards' concept for model and dataset documentation proposed by a number of prominent researchers that included ex-Googler and current 'Hugger' Margaret Mitchell).[33]

After a successful 100 million USD funding round in 2022, Hugging Face received more than 200 million USD in additional funding in fall 2023, with investment from core industry players like Google, Amazon, Nvidia, Intel, AMD, Qualcomm, IBM, and Salesforce, bringing its valuation to about 4.5 billion USD.[34] The firm is developing a business model where it can bundle additional 'premium' deployment features, lowering the barrier to entry for less technical users or those simply seeking convenience. For instance, one product, 'Inference Endpoints,' advertises itself as a way to 'deploy models in minutes' on Hugging Face's own infrastructure. Similarly, their 'AutoTrain' product allows one to 'train, evaluate and deploy state-of-the-art Machine Learning models' by simply uploading a dataset, without having to write a single line of code.[35] These premium, public- or small-enterprise-oriented offerings harness the platform's network effects (as now *the* home for many open models) to add additional extra features or services – making money through classic transaction fee-oriented intermediation. Hugging Face also allows the deployment of models on its platform in web applet 'spaces' that can be easily shared publicly and accessed by others, charging hourly for the hardware these run upon.

---

[29]Lewis Tunstall, Leandro von Werra, and Thomas Wolf, *Natural Language Processing with Transformers* (O'Reilly Media, Inc, 2022) xii.
[30]Ibid xii.
[31]Alexandra Sasha Luccioni and others, 'Stable Bias: Analyzing Societal Representations in Diffusion Models' (arXiv, March 2023) <https://doi.org/10.48550/arXiv.2303.11408>.
[32]Melissa Heikkilä, 'Inside a Radical New Project to Democratize AI' (*MIT Technology Review*, July 2022) <www.technologyreview.com/2022/07/12/1055817/inside-a-radical-new-project-to-democratize-ai/> accessed 30 September 2023.
[33]See Margaret Mitchell and others, 'Model Cards for Model Reporting' in *Proceedings of the 2nd ACM Conference on Fairness, Accountability and Transparency* (FAT* '19, ACM, 2019) <https://doi.org/10/gftgjg>; used in <https://huggingface.co/docs/hub/model-cards>.
[34]Kyle Wiggers, 'Hugging face raises $235M from investors, including salesforce and Nvidia' (*TechCrunch*, August 2023) <https://techcrunch.com/2023/08/24/hugging-face-raises-235m-from-investors-including-salesforce-and-nvidia/> accessed 20 September 2023.

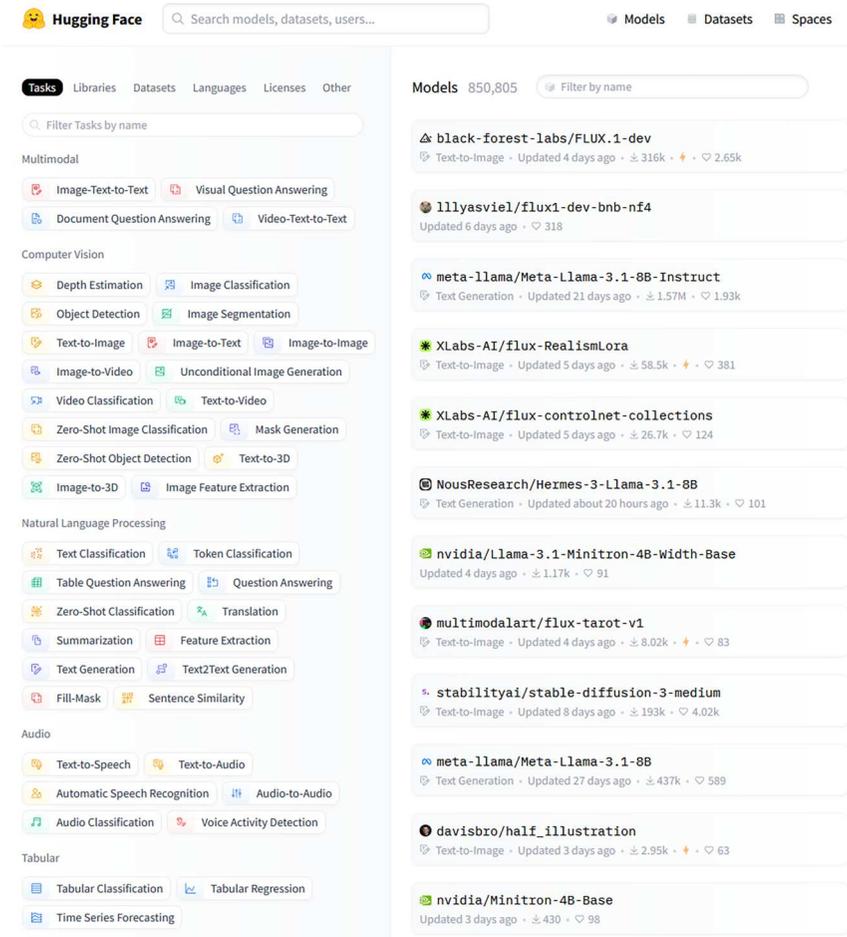

**Figure 1.** A screenshot of the Hugging Face platform, showing models 'trending' in August 2024.

---

[35]Quotes from Hugging Face publicity materials; see archived versions as of August 2023 at <https://web.arch ive.org/web/*/huggingface.co>.

### 2.1.1. Features of model marketplaces

A range of platforms exist that we believe can best be understood as model marketplaces (Table 1). Some of these might be termed *general-purpose model marketplaces* and involve a wide range of AI development functionality. Hugging Face is the best-known example, where individuals and organisations can access, modify, and deploy a range of models designed for tasks ranging from translation and text-generation to image-recognition and classification. Replicate, a start-up from Berkeley, California, operates a similar, although less popular service, allowing users to run a range of open-source models in the cloud through their web interface. Similar offerings have been developed by firms including gravityAI and Modelplace.AI.

As software, models can also be hosted on generic *software development platforms*. These services, such as GitLab or GitHub, may not have been created with machine learning research in mind, and have yet to integrate their repositories directly into the industry leading AI deployment platforms in the way that Hugging Face has. Nevertheless, they still can – and are – be used to host datasets and models, and may be used as part of an MLOps pipeline by certain actors.[36] These platforms could theoretically incorporate more features oriented specifically towards hosting and sharing models in the future if they see it in their business interest to do so – and have worked with organisations including Hugging Face on policy initiatives promoting open-source AI.[37] The proliferation of powerful image generation models has also led to more niche, *image-generation model marketplaces*. These provide a platform where users can upload and download tuned versions of models like Stable Diffusion XL or Stability AI's Control-LoRA, as well as special training data, easy access to (and support with) model prompting and 'prompt engineering,' and social features such as blogs, instructional forums, and 'community challenges.' Examples of this kind of service targeted towards amateur members of the public involve Civitai, PixAI, and Mage.Space; others seek to target more professional content creators. (Note here that other well-known image generation tools, such as Midjourney, allow one to query their models but not easily download or upload their own modifications; we consider third-party model upload to be the key feature of a model marketplace.)[38]

---

[36]For instance to facilitate 'continuous integration and deployment,' see Kreuzberger, Kühl, and Hirschl (n 25).
[37]Peter Cihon, 'How to Get AI Regulation Right for Open Source' (26 July 2023) <https://github.blog/2023-07-26-how-to-get-ai-regulation-right-for-open-source/> accessed 17 November 2023.

Table 1. Functionalities and characteristics of identified model marketplaces as of November 2023.

| Model Marketplace | Can users Upload models? | Can users Query models? | Can users Query models via API? | Developer Monetisation tools? | Model types | Jurisdiction | Founded (Marketplace features may have come later) | Content Policy | Policy Explicitly beyond output/use? |
|---|---|---|---|---|---|---|---|---|---|
| Hugging Face | Yes | Yes | Yes | No | Generic | USA/France | 2016 | Yes | Yes |
| GitHub | Yes | No | No | No (sponsorship possible) | Generic | USA | 2008 | Yes | Yes |
| Civitai | Yes | Yes ('Diffusion Partners') | No | Yes ('bounties') | Image Generation | USA | 2023 | Yes | No |
| Mage.Space | Yes (via a form) | Yes | Yes (in beta) | Yes ('Creators Program') | Image Generation | USA (Ollano Inc.) | 2022 | Yes | No |
| Replicate | Yes | Yes | Yes | No | Generic | USA | 2019 | Yes | No |
| PixAI | Yes | Yes | No | Yes (platform credits only) | Image Generation | USA (Mewtant, Inc., DE) | 2022 | Yes | No |
| gravityAI | Yes (via form) | Yes | Yes | Yes | Generic | USA | 2020 | Yes (brief) | No |
| Modelplace.AI | Yes (via email) | Yes | Yes | Yes | Computer vision | USA (OpenCV.ai Corp.) | 2020 | No | No |
| Tensor.art | Yes | Yes | Yes | No | Image Generation | HK | 2023 | Yes | No |
| Kaggle Models | Yes (via form) | No | No | No | Generic | USA (Google LLC) | 2010 | Yes | No |
| Shoggoth Systems | Yes | No | No | No | Generic | Unknown (peer-to-peer) | 2023 | No | No |

## 2.2. Moderating models as a difficult policy problem

Models have a range of characteristics that make them hard to moderate, even for platforms that wish to act rigorously and in good faith with some notion of public safety or responsible behaviour in mind. In this section, we provide an overview of two sources of this difficulty: the unique technical characteristics that models exhibit as software artefacts, and the current lack of clarity regarding the legal liability that model marketplaces have as important actors in AI 'value chains.'[39]

### 2.2.1. Content inside models

AI models are unusual forms of software insofar as some can be thought of as 'containing' other content. Science fiction author Ted Chiang neatly called large language models 'blurry JPEGs of the Web,' drawing analogies to compression.[40] Information about the world can be retrieved from these models. Such information is content, and such content in turn raises traditional content moderation challenges.

AI models, particularly generative systems, have been shown to **memorise** datasets. Memorisation in language models can be tested by prompting the model with a phrase that is in the training dataset, and seeing if, and how often, it returns words that follow that prompt at least once in the training data. GPT-J, an open-source language model similar to GPT-3, has been shown to memorise at least 1% of its training data.[41] Some researchers have argued memorisation can be mitigated with de-duplication of the training data,[42] or through strategic fine-tuning.[43] Selective differential privacy might help protect certain categories of data within models, such as phone numbers or social security numbers, which previous studies have shown to be significantly at risk.[44] As it stands today however, portions of training data can be reconstructed from models. Such 'model inversion' attacks present varied legal and societal risks and have been noted in the data protection community for many years in relation to a wider array of machine learning models,[45] but no longer require additional datasets or special know-how – today, one could feasibly conduct such 'attacks' on conversational LLMs by deploying well-designed prompts.

---

[38] These image-generation services typically are closer to classic 'AI-as-a-service' business models. Cobbe and Singh (n 9), serving a closed set of models on closed infrastructure. Some services focus on proprietary models (e.g. OpenAI and the DALL-E model series; Midjourney), while others offer a closed but constantly changing set of open source models (e.g. Invoke AI, Wombo Dream, Night Café). Some of these services are also in flux, with indications that they are moving towards a more open marketplace model. For example, OpenAI's 'GPT Store', which allows custom 'system prompts' to create different flavours of chatbot, although not yet full finetuning. However, as these platforms do not allow users to upload models (yet), we exclude them from analysis in this paper.
[39] Jennifer Cobbe, Michael Veale, and Jatinder Singh, 'Understanding Accountability in Algorithmic Supply Chains' in *Proceedings of the 2023 ACM Conference on Fairness, Accountability, and Transparency* (FAccT '23, ACM, June 2023) <https://doi.org/10.1145/3593013.3594073>.
[40] Ted Chiang, 'ChatGPT is a blurry JPEG of the web' (*The New Yorker*, February 2023) <www.newyorker.com/tech/annals-of-technology/chatgpt-is-a-blurry-jpeg-of-the-web> accessed 14 September 2023.
[41] Nicholas Carlini and others, 'Quantifying Memorization Across Neural Language Models' (ICLR 2023, 2023) <https://doi.org/10.48550/arXiv.2202.07646> accessed 4 July 2023.
[42] Nikhil Kandpal, Eric Wallace, and Colin Raffel, 'Deduplicating Training Data Mitigates Privacy Risks in Language Models' in *Proceedings of the 39th International Conference on Machine Learning* (ICML '22, June 2022) <https://proceedings.mlr.press/v162/kandpal22a.html> accessed 5 July 2023.
[43] Ronen Eldan and Mark Russinovich, 'Who's Harry Potter? Approximate Unlearning in LLMs' (arXiv, October 2023) <https://doi.org/10.48550/arXiv.2310.02238> accessed 30 October 2023.

Even if memorisation may one day be mitigated via technical means, even more difficult to tackle are issues relating to the **semantic reconstruction** of information. We use this to refer to information which may not be in exactly the same form as the input data, but expresses the same ideas or concepts. For example: a visual generative system might reproduce art of a certain exact style; an AI text-to-audio system might reproduce someone's voice but without any of the words they are 'speaking' actually having occurred in a training dataset; and a text-generation system might list a biography of somebody which combines facts inferred from multiple sources.

Many studies consider language models as knowledge-bases, and look to extract structured knowledge from them about entities such as living persons, synthesising data automatically from many sources without manual entity connection or resolution.[46] This composite nature of information that underlies statistical LLM outputs makes their moderation inherently difficult.[47] As the underlying data that fed a model is not necessarily visible to the operators of a model marketplace, concerning or liability-attracting content inside models may not become evident for platform operators until the model is already being deployed and used widely by users. Even were such data to be visible, semantic reconstruction means these problems may only be discoverable after a machine learning system has connected the dots between many disparate data points.

### 2.2.2. Liability for content inside models

The challenge here is a wicked one, as even for a good faith organisation attempting to build safe and responsible model marketplaces there often may be no obvious way to identify the potential for memorisation or semantic reconstruction, or to distinguish between permissible and impermissible forms of either.

---

[44] Carlini and others (n 41); Weiyan Shi and others, 'Selective Differential Privacy for Language Modeling' in *Proceedings of the 2022 Conference of the North American Chapter of the Association for Computational Linguistics: Human Language Technologies* (Association for Computational Linguistics 2022).

[45] Michael Veale, Reuben Binns, and Lilian Edwards, 'Algorithms That Remember: Model Inversion Attacks and Data Protection Law' (2018) 376 *Philosophical Transactions of the Royal Society A* 20180083 <https://doi.org/10/gfc63m>.

[46] Badr AlKhamissi and others, 'A Review on Language Models as Knowledge Bases' (arXiv, April 2022) <https://doi.org/10.48550/arXiv.2204.06031> accessed 5 October 2023.

[47] David Glukhov and others, 'LLM Censorship: A Machine Learning Challenge or a Computer Security Problem?' (arXiv, July 2023) <https://doi.org/10.48550/arXiv.2307.10719> accessed 15 September 2023.

On one hand, the memorisation of content, if relatively easily accessible from the model, could quite uncontroversially could be understood as functionally the same as hosting that content. It seems unlikely that many would argue a compressed file, where the method of decompression is common knowledge, should not be treated the same in legal or policy terms as the file itself. Doing otherwise would allow the easy transfer or communication of information, such as child sexual abuse material or copyrighted content, without civil or criminal liability. If a text-generation model is trained on a corpus of data that contains illegal, sensitive personal, or copyright infringing material, and can memorise it and reconstruct it when prompted under conditions that meet some threshold of simplicity, than the logical corollary is that the entities hosting the model may face legal liability if intermediary liability conditions, typically a failure of notice-and-takedown, are met.

Semantic reconstruction is more legally ambiguous and regime-dependent. There is no general answer as to whether reconstructed content, which may for example replicate the style of an artist or reproduce protected characters, would violate copyright or other relevant intellectual property law. The gradient between 'substantially similar' and 'distinguishably different' content in the context of AI is effectively still to be determined by courts.[48]

While facts cannot be copyrighted, they may be subject to other legal regimes. European data protection law (as well as many very similar domestic regimes internationally) requires a legal basis and safeguards to process certain facts, as its scope encompasses all information that 'relates' to an identifiable natural person by means of content, purpose or effect, even including opinions, regardless of their accuracy.[49] Where such information is 'special category,' such as the political opinions of an individual, they receive heightened protection, regardless of accuracy.[50] It is not the case that privacy cannot extend to public spaces, at least in terms of European human rights law, particularly in cases where structured, queryable information analogous to a dossier is being gathered on a person.[51] Significant related case-law in many jurisdictions concerns the interaction of privacy and one of the great structuring forces of online information – search engines – typically captured under the term the 'right to be forgotten.' While individuals might have few expectations of privacy in a language-model-derived synthesis of all the work-related bios they have placed online, they may feel differently if a language model chooses to synthesise the bios with information from identified posts on public social media pages, particularly if they cross contexts, such as material dating from their university years, posted in another language, or intended for a specific audience, such as friends in an online queer community.[52]

---

[48] Pamela Samuelson, 'Generative AI Meets Copyright' (2023) 381(6654) *Science* 158 <https://doi.org/10.1126/science.adi0656>; Katherine Lee, AFeder Cooper, and James Grimmelmann, 'Talkin' 'Bout AI Generation: Copyright and the Generative-AI Supply Chain' (arXiv, September 2023) <https://doi.org/10.48550/arXiv.2309.08133> accessed 4 October 2023.

[49] Nadezhda Purtova, 'The Law of Everything. Broad Concept of Personal Data and Future of EU Data Protection Law' (2018) 10(1) *Law, Innovation and Technology* 40 <https://doi.org/10/gd4rmh>.

[50] Regulation (EU) 2016/679 of the European Parliament and of the Council of 27 April 2016 on the protection of natural persons with regard to the processing of personal data and on the free movement of such data, and repealing Directive 95/46/EC (General Data Protection Regulation) [2016] OJ L119/1 (GDPR) art 9; Case C-252/21 *Meta Platforms and Others* ECLI:EU:C:2023:537, para 69.

But in politicised and securitised policy domains like terrorism, many jurisdictions have designed extremely broad frameworks with an intentionally low liability threshold which explicitly include tools that can create specific pieces of content. For instance, counterterrorism law in the UK seeks to criminalise the dissemination of 'information of a kind likely to be useful to a person committing or preparing an act of terrorism.'[53] A language model that could rephrase the gist of *The Anarchist Cookbook* or other training manuals would likely meet this broad minimum standard. UK law covering child sexual abuse material is even more explicit in this regard, as it specifically covers 'pseudo-photographs:' potentially computer-generated images which appear to be indecent photographs and which convey the impression that the person shown is a child. The UK framework specifically states that the definition of pseudo-photographs include 'data [..] capable of conversion into an indecent pseudo photograph.'[54] There are undoubtedly conditions where a generative model would fall into that category.

Other extant legislation in some policy domains does explicitly outlaw tools, but does seek to govern computer-generated images which possess certain characteristics. England and Wales have laws against intimate image abuse, where liability can occur if an image 'appears' to show an intimate scene featuring the victim.[55] In that case, non-consensual AI generated images of individuals could very well meet this existing standard. The question then arises as to what point a model is considered legally equivalent to an image that could be produced by it. When it is bundled with a query, or a prompts that can produce this effect are publicised? Or simply when it has the latent ability to produce such an image of an existing, natural person, without external imagery being provided?

---

[51] Lilian Edwards and Lachlan Urquhart, 'Privacy in Public Spaces: What Expectations of Privacy Do We Have in Social Media Intelligence?' (2016) 24(3) *International Journal of Law and Information Technology* 279 <https://doi.org/10/gfzqk9>.

[52] Helen Nissenbaum, *Privacy in Context: Technology, Policy, and the Integrity of Social Life* (Stanford University Press, 2010); Anthony Henry Triggs, Kristian Møller, and Christina Neumayer, 'Context Collapse and Anonymity among Queer Reddit Users' (2021) 23(1) *New Media & Society* 5 <https://doi.org/10.1177/1461444819890353>.

[53] Terrorism Act 2000, s 58(1)(a).

[54] Protection of Children Act 1978, s 7.

[55] Sexual Offences Act 2003, s 66B.

### 2.2.3. Models as tools

As discussed, some laws regulating the most extreme information-related offences, such as terrorism and child sexual abuse material (CSAM), already explicitly collapse 'content' and the tools generating it into the same category of information under the law. However, models are tools which can be generally used for many tasks beyond the creation of specific highly illegal content, and this creates a significant moderation challenge for the platforms that host and facilitate access to them.

How model marketplace trust and safety teams might best consider these tasks and uses is not clear. One option would be to consider *intended* uses. Ideally those would be provided by developers on a model card accompanying the artefact,[56] but they could also be inferred from the model's title and description. Another would be for the platform to intervene following *realised* misuses, after gaining knowledge of a model being misused, regardless of the developer's intent or the previously analysed perceived capabilities of the model. Finally, moderators could consider *potential* or *likely* uses – relating to the capabilities of the model and the relative societal risks. In legal terms, these three valences resemble *intent*, *knowledge*, and *strict liability requirements*, respectively.

From a trust and safety perspective, model marketplaces can most easily consider the intended use of models posted to their platforms – but understanding either potential or realised use (already mandated by some existing legal frameworks, such as the potential generation of terrorism/CSAM content described in Section 2.2.1) is resource intensive, requiring foresight capacity, real-world monitoring, or the ability to critically appraise evidence about the impact of systems on the world. There are some parallels here to the highest-stakes and most difficult decisions being made in the social media content policy realm – for example relating to the removal of the accounts of major political figures in a context of potential 'off platform' political violence – but these are high-stakes exceptions to the relatively rote norm of content moderation at scale.

In terms of the risks that models might pose as tools, we signpost readers to other work rather than exhaustively elaborating here.[57] Misuse of software is difficult to anticipate because of its generative nature, allowing it to be leveraged across many tasks, even ones their creators did not envisage[58] – but the social and political stakes are certainly high.[59]

---

[56]Mitchell and others (n 33).
[57]See generally Weidinger and others (n 15).

## 2.3. Model marketplaces and existing regulation

AI intermediaries are not yet playing a major role in today's high-level international AI policy discussions. However, given their structural position in the AI value chain, it seems very likely they will soon have to. A previous generation of internet hosting intermediaries over time became deeply enmeshed in law and policy as the salience of content-related issues grew: the history of platform governance in the user-generated content space demonstrates that legal pressure (e.g. from copyright holders), commercial pressure (e.g. advertiser sensibilities), and policy scrutiny (from civil society and powerful government actors) are generally the primary drivers of meaningful changes in industry moderation practices.[60] European officials have, for instance, since the mid-2000s been active in seeking to use informal negotiation and policy fora (such as the EU Internet Forum) to pressure social media platforms to invest more resources in content detection and removal processes relating to child safety and violent extremism.[61] Model marketplaces are new actors in a complex ecosystem and have yet to receive such sustained attention from regulators or pressure groups.

The landscape of formal, binding AI governance is still inchoate.[62] Model marketplaces currently sit in either a policy vacuum (minus generic intermediary liability shields) or one where they may be potentially subject to regimes either for extreme content or social media regulation which do not consider their unique characteristics and challenges (as explored in Section 2.2.2 above). The leading platform regulation and AI policy frameworks currently being debated, implemented and interpreted – in particular the EU's Digital Services Act (DSA) and AI Act – generally fail to capture the important role of model marketplaces as either high-stakes user-generated content platforms or more generally as critical actors in AI supply chains. The DSA is oriented towards traditional 'social media' content hosts, with its main provisions applying only to 'very large online platforms' (VLOPs) with a 45 million monthly active user threshold.[63] Unlike social media and other user-generated content platforms, models on marketplaces indirectly affect individuals and environments through a much smaller number of user-developers, and so marketplaces tend to fall out of scale thresholds in existing platform law. The enormous classic software repository, GitHub, has not been designated as a VLOP, claiming only 11-12 m EU monthly active users in August 2023.[64] Hugging Face does not disclose similar numbers, but is likely to be significantly smaller.[65] Even where platform regulation does apply, it was built with traditional user-generated content in mind, not the specific challenges of *powerful AI tools*.

---

[58]Jonathan L Zittrain, 'The Generative Internet' (2006) 119 *Harvard Law Review* 1974.

[59]Complicating things further, AI models evoke parallels to other 'dual-use' tools, such as drones, lasers, 3D printers, nuclear components, and certain chemicals, which have clear positive and negative uses. There is a difference however – the riskiest of these dual-use technologies are typically produced in moderately or highly-regulated industries, in sharp contrast to today's model marketplaces, where anyone can make an account and upload and download powerful models with virtually no policy constraints.

[60]Jillian C York, *Silicon Values: The Future of Free Speech Under Surveillance Capitalism* (Verso, 2021); Robert Gorwa, *The Politics of Platform Regulation: How Governments Shape Online Content Moderation* (Oxford University Press, 2024).

[61]Christopher T Marsden, *Internet Co-Regulation: European Law, Regulatory Governance and Legitimacy in Cyberspace* (Cambridge University Press, 2011); Robert Gorwa, 'The Platform Governance Triangle: Conceptualising the Informal Regulation of Online Content' (2019) 8(2) *Internet Policy Review* 1 <https://doi.org/10.14763/2019.2.1407>.

[62]Michael Veale, Kira Matus, and Robert Gorwa, 'AI and Global Governance: Modalities, Rationales, Tensions' (2023) 19(1) *Annual Review of Law and Social Science* 255 <https://doi.org/10.1146/annurev-lawsocsci-020223-040749>.

In terms of the EU AI Act, scholars have noted that the framework is overly 'focused on the model itself rather than the use-case specific application.'[66] Edwards has noted that, for this reason, the AI Act has an extremely limited view of the multiple actors involved in the inception, training, tuning, and deployment of powerful machine learning models.[67] The companies running model marketplaces themselves have further protested at the proposed instrument's lack of understanding of the open-source software development system.[68] The AI Act seeks to exempt models with open licenses from obligations unless they are put into service for high risk purposes, although under what conditions 'putting into service' is triggered by uploading a model designed for a certain purpose seems extremely unclear.[69]

---

[63] Regulation (EU) 2022/2065 of the European Parliament and of the Council of 19 October 2022 on a Single Market For Digital Services and amending Directive 2000/31/EC [2022] OJ L265/1 (DSA) art 33(a).
[64] See disclosed DSA statistics at <https://github.com/github/transparency/tree/main/data/eu_dsa>.
[65] A lot depends on the methodology. Hugging Face's most downloaded repositories report between 60 and 40 million 'downloads last month.' It is unclear whether these are cumulative numbers or numbers that refer to monthly API calls. If the latter is the case, given that basic web traffic metrics indicate that huggingface.co receives about 18 million visits a month, (Similarweb, 'Huggingface.Co Traffic Analytics, Ranking Stats & Tech Stack' (*Similarweb*, 2023) <https://perma.cc/GSW4-YN3X> accessed 30 October 2023) the platform would currently fall beneath the DSA VLOP threshold if it implemented an account-based system or other measures to link these instances of 'model use' to distinct users. See figures at <https://huggingface.co/models>. That said, the non-VLOP provisions of the DSA, such as requirements to proportionally enforce terms and conditions, would still apply.
[66] Philipp Hacker, Andreas Engel, and Marco Mauer, 'Regulating ChatGPT and Other Large Generative AI Models' in *Proceedings of the 2023 ACM Conference on Fairness, Accountability, and Transparency* (FAccT '23, ACM, June 2023) <https://doi.org/10.1145/3593013.3594067> 1115.
[67] Lilian Edwards, 'Regulating AI in Europe: Four Problems and Four Solutions' (Ada Lovelace Institute, London, UK, 2022) <https://perma.cc/E9S4-W8LT> accessed 20 April 2023.
[68] Cihon (n 37).
[69] Regulation (EU) 2024/1689 of the European Parliament and of the Council of 13 June 2024 laying down harmonised rules on artificial intelligence and amending Regulations (EC) No 300/2008, (EU) No 167/

It should not be unsurprising then that today's leading AI regulation does not conceive of model marketplaces or other increasingly important AI intermediaries that might host and provide access to AI systems as important regulatory targets. The AI Act would not see model marketplaces as either model developers or 'users,' although the more generic powers in the Market Surveillance Regulation that will apply to AI regulators under the AI Act will give powers to order takedowns and obliged co-operation in harm reduction by intermediaries.[70] Marketplaces might ostensibly be 'distributors' under the proposed framework,[71] but extending the definition of distributor in product safety to online marketplaces would set a head-on collision with intermediary liability law around user-uploaded content – which on online platforms, also includes products.[72] If the AI Act or other future regulatory frameworks eventually develop a more nuanced understanding of the different actors in the AI value chain,[73] they could explicitly seek to design rules for model marketplaces and other AI intermediaries, seeking to implement special standards and obligations around complaints handling, due process, and transparency.

## 3. Governance by model marketplaces: case studies

To illustrate some of the challenges of moderating model marketplaces, it is worth looking at some of the actual cases that they have dealt with so far. In the following section we present short content policy vignettes from 3 different platforms – Hugging Face, GitHub, and Civitai – which are exemplary as the largest general-purpose marketplace, the largest software development platform, and perhaps the largest and fastest growing specialised image-generation model marketplace.

### 3.1. Hugging Face

When Hugging Face launched the service that is now at the core of their business, they initially had no public Terms of Service (ToS), no community guidelines, and no content policy. Their first ToS document, dated 31 May 2021, said very little about moderating the content that the platform hosted. It noted only that:

---

2013, (EU) No 168/2013, (EU) 2018/858, (EU) 2018/1139 and (EU) 2019/2144 and Directives 2014/90/ EU, (EU) 2016/797 and (EU) 2020/1828 (Artificial Intelligence Act) [2024] OJ L.
[70]Regulation (EU) 2019/1020 of the European Parliament and of the Council of 20 June 2019 on market surveillance and compliance of products and amending Directive 2004/42/EC and Regulations (EC) No 765/2008 and (EU) No 305/2011 [2019] OJ L169/1 (Market Surveillance Regulation) arts 7, 14(3)(k).
[71]Commission, 'Proposal for a Regulation of the Parliament and of the Council Laying down Harmonised Rules on Artificial Intelligence (Artificial Intelligence Act) and Amending Certain Union Legislative Acts' COM(2021) 206 final, arts 3(7), 27.
[72]For example, on eBay; see e.g. Case C-324/09 *L'Oréal SA and Others v eBay International AG and Others*. ECLI:EU:C:2011:474.
[73]Cobbe, Veale, and Singh (n 39); Hacker, Engel, and Mauer (n 66).

> Your Content must not be misleading or unlawful, and must not violate any of these Terms, applicable law and regulation, or infringe or misappropriate any rights of any person or entity. We may remove your Content at any time, at our sole discretion, if we have a concern about your Content.[74]

Following the relatively high-profile content moderation controversy that was the GPT-4chan episode (Section 1), the company appears to have invested some time and resources into creating a broader and more detailed set of platform rules.

### 3.1.1. Constructing a content policy

The new content policy of August 2022 distinguished between 'technical' and 'human' content. While 'human' content such as comments or discussions had some clear guidelines, 'technical' content such as datasets and models were instead subject to a 'public discussion' asking for 'feedback,' but with no publicly articulated principles or policies which these models would be judged against.[75] As the platform grew in popularity, and was increasingly relied upon to share the big generative models being released by leading labs, the policy saw a major change in June 2023. This change removed the prior distinction Hugging Face had made between 'human' and 'technical' forms of content and provided a list of characteristics that would could lead to all types of content being 'restricted,' as well a more process-based set of considerations that might lead to content being 'moderated,' such as through access restrictions.[76] In this new policy, Hugging Face staff stated that they would now pay attention to the 'origin of the ML artifact, how the ML artifact is handled by its developers, and how the ML artifact has been used.'[77]

In the restricted content list (Figure 2), we can see a range of broad categories and different types of moderation rationales. Some involve intent-based classifications, such as content 'designed' or 'created for' certain ends. Some are based on realised consequences, such as content that 'harms others' or is 'used [..] for' other ends. Many categories use the term 'promotes,' which seems to encompass both observations of the purpose of the model and considerations of its use in practice. The policy also features some terminology borrowed from the classic realm of user-generated content platform governance, such as Facebook's infamous 'coordinated inauthentic behaviour' term. It

---

[74]Hugging Face, 'Terms of Service' (June 2021) <https://web.archive.org/web/20210622075735/>; <https://huggingface.co/terms-of-service> accessed 17 September 2023.
[75]Hugging Face, 'Content Policy' (August 2022) <https://web.archive.org/web/20221130213223/>; <https://hug gingface.co/content-guidelines> accessed 16 September 2023.
[76]Hugging Face, 'Content Policy' (July 2023) <https://web.archive.org/web/20230717150419/>; <https://huggingface.co/contentguidelines> accessed 16 September 2023. A further change occurred in August 2023 but brought no substantial changes relevant to this paper.
[77]Ibid.

- *Unlawful, defamatory, fraudulent or intentionally deceptive Content, including, but not limited to co-ordinated or other inauthentic behavior, disinformation, phishing or scams;*
- *Content that harms others;*
- *Content promoting discrimination (see our Code of Conduct), or hate speech;*
- *Content harassing, demeaning, or bullying;*
- *Sexual content used or created for harassment, bullying, or without explicit consent of the people represented;*
- *All sexual content involving minors;*
- *Content that promotes or glorifies violence or the suffering or humiliation of another;*
- *Content that promotes or induces unlawful or fraudulent currencies, securities, investments, or other transactions;*
- *Content published without the explicit consent of the people represented;*
- *Spam, such as advertising a product or service, or excessive bulk activity;*
- *Cryptomining practices;*
- *Content that infringes or violates any rights of a third party or an applicable License;*
- *Content that violates the privacy of a third party;*
- *Content that violates any applicable law or regulation;*
- *Content that attempts to transmit or generate code that is designed to disrupt, damage or gain unauthorized access to a computer system or device;*
- *Content that is malware, a trojan horse or virus, or other malicious code;*
- *Proxies that are primarily designed to bypass restrictions imposed by the original service provider;*
- *Content that promotes high-risk activities, including but not limited to, weapons development, self-harm, suicide, gambling, plagiarism, scams or pseudo-pharmaceuticals.*

**Figure 2.** Hugging face restricted content policy (as of August 2023).

gives the company very broad latitude to moderate, including content that violates 'any applicable law or regulation' (seemingly in any and all jurisdictions), that violates the privacy of any third party, and importantly, violates an 'applicable license' – a subject to which we now turn.

### 3.1.2. Enforcing Licenses – or, Xi Jinping won't sing

AI systems are increasingly being released under increasingly complex and atypical software licenses. Software licenses slowly emerged from the 1960s onwards, gaining traction once software fell more firmly under copyright protection in the 1980s.[78] As a legal tool, licenses seek to make the reuse of intellectual property conditional on the adherence to certain conditions. They have long been used in an attempt to govern the downstream uses of generative technologies. For instance, open-source advocates developed 'copyleft' licenses that sought to permit others to build upon or modify software without cost, but only if they did not later release the results under proprietary/for-profit licenses of their own.[79] Other licenses included distribution-related conditions, such as those seeking to prevent military uses of their software,[80] broad conditions such as the JSON license's statement that '[t]he Software shall be used for Good, not Evil,'[81] or satirical licenses, such as the 'Anyone But Richard M Stallman' (ABRMS) license, which provides that individuals can do whatever they want with the software, unless they are Stallman, the controversial developer of the copyleft GNU General Public License.[82]

---

[78]WS Humphrey, 'Software Unbundling: A Personal Perspective' (2002) 24(1) *IEEE Annals of the History of Computing* 59 <https://doi.org/10.1109/85.988582>; Amy Thomas, 'The First Software Licensing Agreement and Its Relationship with Copyright Law' (*CREATe*, October 2019) <www.create.ac.uk/blog/2018/11/14/the-first-software-licensing-agreement-and-its-relationship-with-copyright-law/> accessed 6 October 2023.

In the ongoing (somewhat) 'open-source' AI boom,[83] licensing AI models upon release has become common. These licenses now often contain ambitious conditions that go far beyond distribution.[84] We do not weigh in on exactly when such restrictions mean an license is not 'open-source,' as the use and alleged abuse of the term remains a subject of ongoing controversy. But Hugging Face has been especially vocal in their support of the OpenRAIL family of licenses, seeing them as a promising method of transmitting and enforcing norms in the ML community.[85] The OpenRAIL license, which can be applied to several points in the AI development cycle, states that the license is revoked if the licensee or any third party under their control uses the tool for certain purposes, such as to infer certain categories of sensitive data about an individual, predict health characteristics for the purposes of insurance pricing, attempt to predict criminality, or synthesise undeclared realistic representations of people or events.[86] Furthermore, as indicated explicitly in their content policy, Hugging Face will restrict '[c]ontent that infringes or violates [..] an applicable License' (Figure 2). Furthermore, Hugging Face offers model uploaders the possibility to not just upload a license in their repository (as GitHub does) but to oblige users to explicitly agree to it, and to provide their contact details to the repository owners, before accessing the model (Figure 3). This step helps ensure that contracts and agreements do indeed bind the user, as passive 'browsewrap' contracts are generally seen to be weaker when compared to explicit 'clickwrap' contracts, although the latter too may have issues in some jurisdictions.[87]

---

[79]Richard M Stallman, *Free Software, Free Society* (Free Software Foundation, 2015).
[80]Steve Dierker and Volker Roth, 'Can Software Licenses Contribute to Cyberarms Control?' in *Proceedings of the New Security Paradigms Workshop* (NSPW '18, ACM, August 2018) <https://doi.org/10.1145/3285002.3285009>.
[81]JSON, 'The JSON License' (2002) <www.json.org/license.html> accessed 7 October 2023.
[82]Landon Dyer, 'Another Assembler' (*Dadhacker*, February 2014) <https://web.archive.org/web/20140207084017/>; <www.dadhacker.com/blog/?p=2106> accessed 7 October 2023.
[83]Widder, West, and Whittaker (n 10).
[84]Danish Contractor and others, 'Behavioral Use Licensing for Responsible AI' in *2022 ACM Conference on Fairness, Accountability, and Transparency* (FAccT '22', ACM, 2022) <https://doi.org/10.1145/3531146.3533143>; Veale, Matus, and Gorwa (n 62).
[85]Giada Pistilli and others, 'Stronger Together: On the Articulation of Ethical Charters, Legal Tools, and Technical Documentation in ML' in *Proceedings of the 2023 ACM Conference on Fairness, Accountability, and Transparency* (FAccT '23, ACM, 2023) <https://doi.org/10.1145/3593013.3594002> accessed 14 June 2023.
[86]'Responsible Artificial Intelligence Source Code License' (*Responsible AI Licenses (RAIL)*, November 2022) <www.licenses.ai/source-code-license> accessed 9 November 2023.

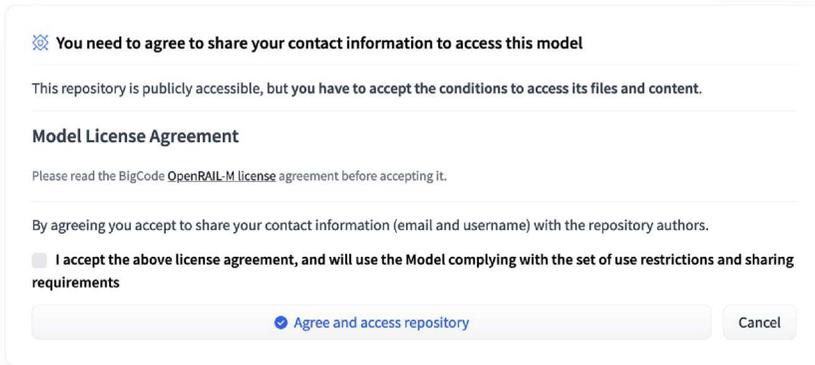

**Figure 3.** Dialogue box from the BigCode starcoder model, asking users to explicitly accept license terms and forward contact details from users' accounts.

However, OpenRAIL is not the only license in town. Popular models on Hugging Face are licensed in a wide variety of ways that limit the behaviour of users.[88] For example, Falcon 180bn, (a large language model created by the Technology Innovation Institute in Abu Dhabi) is available for download and deployment on Hugging Face with a license which stipulates that their model cannot be used '[f]or the purpose of exploiting, harming or attempting to exploit or harm minors and/or living beings in any way.'[89] Baichuan 2, a leading Chinese large language model, has a license only available in Chinese which states that users cannot not violate the laws of the People's Republic of China.[90] Meta's LLaMA 2 Community License is withdrawn if individuals use Meta's language model to train or improve any other model, or if the individual initiates a copyright claim pertaining to LLaMA against Meta.[91]

---

[87]Clifford Fisher and others, 'Evolution of Clickwrap & Browsewrap Contracts' (2022) 48(2) *Rutgers Computer and Technology Law Journal* 147; Susan Corbett, 'Computer Game Licences: The EULA and Its Discontents' (2019) 35(4) *Computer Law & Security Review* 453 <https://doi.org/10.1016/j.clsr.2019.03.007>.
[88]In late September 2022, 0.54% of Hugging Face model repositories made use of an OpenRAIL license. By the end of January 2023, that proportion had risen to 9.81%. (Paul Keller and Nicolò Bonato, 'Growth of Responsible AI Licensing. Analysis of License Use for ML Models' (*Open Future*, Open Future Foundation February, 2023) <https://openfuture.pubpub.org/pub/growth-of-responsible-ai-licensing/release/2> accessed 29 August 2023) See ibid for a deeper analysis of licensing trends on HF.
[89]Technology Innovation Institute, 'Falcon 180B TII License Version 1.0' (*Hugging Face*, September 2023) <https://perma.cc/JWX5-3QC4> accessed 21 November 2023.
[90]Baichuan Intelligent Technology, 'Baichuan 2 Model Community License Agreement' (*Hugging Face*, September 2023) <https://perma.cc/WC8H-3PAP> accessed 16 September 2023.
[91]Meta, 'Llama 2 Community License Agreement' (July 2023) <https://ai.meta.com/llama-project/license> accessed 8 October 2023. Meta has added such terms to open source licenses in the past and had to back down from them under developer pressure, in particular in the case of ReactJS,

In both 2022 and 2023, Hugging Face removed models that related to Chinese Premier Xi Jinping. The 2023 removals were of a model that, from the disabled repository, appears to have been a diffusion model based on the SoftVC VITS Singing Voice Conversion framework, which presumably allowed users to produce audio of the leader singing whatever they desired. The first was called 'XiJinPing_Singing' and the second 'WinnieThePoohSVC_sovits4.'[92] Hugging Face has created a repository that purportedly lists all of the government and industry take-down requests they have received (10 as of November 2023),[93] a practice that has long been part of GitHub's government compliance practices. Following best practices for transparency, GitHub publishes the full text of these requests as well; however the Xi Jinping takedown request made to Hugging Face on 22 August 2023 has been partially redacted, and the takedown of the Winnie the Pooh–named repository came with no request at all.[94] The source of the takedown request is unclear. Hugging Face initially redacted the record until media outlets reported on the pre-print of this research paper,[95] after which they partially unredacted the request, claiming that the rightsholder had made it. It remains unclear whether the request was made under official pressure. In any case, the concern we have is with the possibilities these events illustrate.

The take-down conclusions made publicly by Hugging Face staff cites Section 4 of the underlying project's license. This license states:

> Engaging in illegal activities, as well as religious and political activities, is strictly prohibited when using this project. The project developers vehemently oppose the aforementioned activities. If you disagree with this provision, the usage of the project is prohibited.[96]

Hugging Face employees thus authorised the take-down of the fine-tuned Xi singing model, with the justification that the model was clearly being used for political purposes, thus violating a bespoke statement included as custom license for one of the pieces in the model's 'supply chain.' Describing the

---

see Keith Collins, 'Outraged Programmers Stood up to Facebook over Open Source Licensing and Won. Sort Of' (*Quartz*, September 2017) <https://qz.com/1087865/outraged-programmers-stood-up-to-facebook-fb-over-open-source-licensing-and-won-sort-of> accessed 8 October 2023. However, these terms related to patents, which few people hold; the LLaMA 2 terms relate to copyright, which a huge number of people could reasonably claim is infringed in relation to large language models.

[92] See <https://huggingface.co/WitchHuntTV/XiJinPing_Singing> and <https://huggingface.co/spaces/WitchHuntTV/WinnieThePoohSVC_sovits4>.

[93] See <https://huggingface.co/datasets/huggingface-legal/takedown-notices>.

[94] <https://huggingface.co/datasets/huggingface-legal/takedown-notices/blob/main/2023/2023-08-22-Xi-Jinping.md>; see the GitHub takedown repository for a comparison: <https://github.com/github/gov-takedowns>.

[95] Joseph Cox, 'Hugging Face Removes Singing AI Models of Xi Jinping But Not of Biden' (21 November 2023) <www.404media.co/hugging-face-removes-singing-ai-models-of-xi-but-not-of-biden/> accessed 4 April 2024.

[96] <https://huggingface.co/WitchHuntTV/XiJinPing_Singing/discussions/3> (archived at <https://perma.cc/G 985-MGDC>).

bounds of what is considered 'political' has long been a contested point in platform governance,[97] and the model's satirical outputs may have been illegal under Chinese state defamation laws, but it is evident that the model also could be used to produce political expression that not only would be seen as legitimate in many jurisdictions outside of China, but also would be seen as legitimate satire in other frameworks, such as international human rights norms.[98] It is notable that similar Xi Jinping speech synthesisers have been subject to remarkably similar takedown requests on GitHub in 2021 by the Chinese authorities, on the basis that they breach Chinese law rather than a license condition (concerning content that '[harms] national honor and interest'), but GitHub apparently refused to globally remove the relevant repositories as many listed remain accessible from European IP addresses.[99]

Hugging Face does not have a closed list of licenses it will enforce or recognise on its platform, instead explicitly providing functionality for users to see and choose between any license that has ever been used on Hugging Face.[100] The difficulties this seems likely to present have rarely been raised before in the context of other platforms, as IP and copyright have historically been generally driven by the takedown requests of rights-holders or their agents. Hugging Face seems to be taking a proactive, voluntary approach to enforcing licenses either without legal notification by rightsholders, or due to government takedown – and the licenses within scope of this policy are far from straightforward, containing multiple contested concepts.[101]

### 3.2. GitHub

GitHub is another important AI development intermediary, even although it has not to date explicitly introduced features geared towards the machine learning community. It has been around for longer than any other model marketplace, and thus has grappled for some years with difficult challenges posed by older types of software tools with potentially dual-use valences. Their trust and safety history has included numerous instances of seeking to moderate content with large implications for copyright, cybersecurity, and sexual privacy.

---

[97]Vera Sosnovik and Oana Goga, 'Understanding the Complexity of Detecting Political Ads' in *Proceedings of the Web Conference 2021* (WWW '21, ACM, 2021) <https://doi.org/10.1145/3442381.3450049>.
[98]David Kaye, *Speech Police: The Global Struggle to Govern the Internet* (Columbia Global Reports, 2019) <https://doi.org/10.2307/j.ctv1fx4h8v>.
[99]Beijing Network Industry Association, 'Letter to GitHub' (GitHub Government Takedown Repository, January 2021) <https://github.com/github/gov-takedowns/blob/master/China/2021/2021-01-29-BNIA.md> accessed 17 September 2023.
[100]Hugging Face, 'Licenses' (2023) <https://huggingface.co/docs/hub/repositories-licenses> accessed 17 November 2023.
[101]In a traditional user-generated content context, if an uploader was actually a valid licensee, or benefits from an exemption (e.g. fair use in the US), they could submit a complaint (DSA, art 20) or counter-notice (Digital Millennium Copyright Act (DMCA), 17 U.S.C. §512(g) (United States)) and the moderation decision should be reversed. The kind of licenses that typically govern photos or text are comparatively straightforward – such as whether the terms of a stock image license permit posting on social media or not – making these judgements even more difficult in the model marketplace context.

### 3.2.1. Dual-use and fair use

The US Digital Millennium Copyright Act introduced in 1998 an early statutory 'notice and takedown' system for intermediary liability. It also, however, features a lesser known section seeking to govern certain software use, which prohibits offering to the public a technology designed to circumvent a 'technological protection measure' applied to a copyrighted work.[102] In October 2020, GitHub was issued a DMCA take-down request by the Recording Industry Association of America (RIAA) for the *youtube-dl* tool, a popular piece of software enabling individuals to download videos from YouTube and some other services. GitHub initially removed this repository, but after some public outcry, restored access – apparently after receiving legal advice that the allegedly bypassed 'protections' did not '[require] the application of information, or a process or a treatment, with the authority of the copyright owner, to gain access to the work.' [103] Following this argument, *youtube-dl* was simply engaging with general-purpose Web technologies used by YouTube rather than, for example, seeking to break encryption keys held by the copyright owner and deployed in a 'digital rights management' solution.[104]

This incident illustrates the technical and legal resources needed to critically assess claims around software and its functionality, and is a demonstration of the argument commonly espoused by digitally-oriented civil society that these types of takedowns create 'chilling effects' as platforms over-remove content in order to avoid potential liability.[105] Indeed, in response to the youtube-dl drama, GitHub set up a $1m USD defence fund for developers targeted by overzealous DMCA anti-circumvention takedown requests so that they could make successful counter-notices, and lobbied the U.S. government to introduce broader exemptions from the DMCA.[106] Between June 2021 and September 2023, there appear to have

---

[102]17 U.S.C. §1201. These rules have long been claimed to be overbroad for reasons highly linked to the 'dualuse' debate, particularly insofar as they may render it illegal to develop tools which allow individuals to engage in a privileged act of circumvention, such as 'fair use' in the US. See generally Pamela Samuelson, 'Intellectual Property and the Digital Economy: Why the Anti-Circumvention Regulations Need to Be Revised' (1999) 14(2) *Berkeley Technology Law Journal* 519.
[103]Abby Vollmer, 'Standing up for Developers: Youtube-Dl Is Back' (*The GitHub Blog*, November 2020) <https://github.blog/2020-11-16-standing-up-for-developers-youtube-dl-is-back/> accessed 16 July 2023.
[104]Ibid.
[105]Jonathon W Penney, 'Understanding Chilling Effects' (2021) 106 *Minnesota Law Review* 1451.
[106]Ernesto Van der Sar, 'GitHub Reinstates Youtube-DL and Puts $1M in Takedown Defense Fund' (*TorrentFreak*, November 2020) <https://torrentfreak.com/github-reinstates-youtube-dl-and-puts-1m-in-takedown-defense-fund-201116/> accessed 16 July 2023.

been more than forty instances where GitHub had 'offered to connect' repository owners with legal resources to establish an anti-circumvention-related counterclaim before taking it down.[107] These typically related to repositories used to download multimedia from websites, or to those providing tools to modify video games or other applications. Whether this offer was taken up or not is unclear – despite the potential financial assistance, it is worth noting that a DMCA counter notice requires identification of a user who otherwise may have been anonymous – something that individuals may not be willing to do, lest they become the subject of lawsuits more directly.[108]

### 3.2.2. Dual-use and exploit toolkit misuse

Computer security research consistently relies on software that exhibits strong dual-use properties. An integral role of the cybersecurity community is to disclose, experiment with, and publish software exploits and hacking toolkits. For example, GitHub hosts the Metasploit Framework, a toolkit allowing commonly used bugs and weaknesses in computer systems to be systematised, saved, and deployed down the line. This framework contains exploits such as the EternalBlue set of Windows vulnerabilities – purportedly leaked from the US National Security Agency and used to drive the costly global ransomware attack WannaCry[109] – that are available for penetration testers to use in a modular, easily configurable way. While it could be used to hack into vulnerable systems with malicious intent, this software is also commonly deployed by cybersecurity penetration testers as a way of identifying whether IT systems are insecure.[110]

In 2021, GitHub (which, as mentioned above, was acquired by Microsoft in 2018) removed a piece of proof-of-concept code making use of four Microsoft Exchange exploits, known collectively as ProxyLogon, that were already being widely used by notable hacking groups.[111] Publishing such code after a vulnerability has been addressed in a software update is normal security research practice, and indeed part of the social norms and reputation-establishing practices of security researchers.[112]

---

[107]GitHub labels notices with this offer; we ran a search through all notices on <https://github.com/github/dmca/> as of 17 September 2023 to locate notices labelled in this way.
[108]U.S.C. §512(g)(3)(d).
[109]Kristoffer Kjærgaard Christensen and Tobias Liebetrau, 'A New Role for 'the Public'? Exploring Cyber Security Controversies in the Case of WannaCry' (2019) 34(3) *Intelligence and National Security* 395 <https://doi.org/10.1080/02684527.2019.1553704>.
[110]Filip Holik and others, 'Effective Penetration Testing with Metasploit Framework and Methodologies' in *2014 IEEE 15th International Symposium on Computational Intelligence and Informatics* (CINTI '14, November 2014) <https://doi.org/10.1109/CINTI.2014.7028682>.
[111]Dan Goodin, 'Critics Fume after Github Removes Exploit Code for Exchange Vulnerabilities' (*Ars Technica*, March 2021) <https://arstechnica.com/gadgets/2021/03/critics-fume-after-github-removes-exploit-code-for-exchange-vulnerabilities/> accessed 16 July 2023.
[112]David Bozzini, 'How Vulnerabilities Became Commodities. The Political Economy of Ethical Hacking (1990–2020)' (April 2023) <https://hal.science/hal-04068476> accessed 14 October 2023.

GitHub's removal of this code sparked significant controversy in the cybersecurity community. This seems to have been fuelled – at least in part – by concerns about GitHub's relatively new Microsoft ties, but also by concerns that GitHub was adopting an overly restrictive approach to important – yet dual-use – technologies. A draft content policy shared after the incident states that GitHub will remove malware tools 'that are in support of ongoing and active attacks'[113] although the final policy adopted by GitHub after community consultation was narrower, focusing on potential instances when GitHub servers are being used to deliver malware in active attacks.[114] This allowed GitHub to side-step the issue of censorship by adding friction to the automated download or distribution of content without blocking it entirely – a policy choice with parallels to how social media platforms have historically engaged in 'downranking' and visibility limits when faced with certain politically sensitive moderation choices.[115]

GitHub's experience here is relevant as similar issues seem just around the corner for model marketplaces. When certain AI models present public safety concerns, there will be policy stakeholders seeking to get intermediaries to better defend against those issues. This raises the genuine questions of who, and under what conditions, hosts models known to have potential offensive capabilities, even if they are being hosted – as in the case of dual-use software vulnerabilities – for the purpose developing defensive capabilities and security best practices.

### 3.2.3. Dual-use and image-based abuse

'Deepfake' generation software developers have found a home on GitHub for many years.[116] These tools involve new and improved techniques for media synthesis, which allow one to modify, splice together, and otherwise generate realistic looking hybrid video.[117] One of the earliest controversial repositories, *deepfakes_faceswap*, is a tool which can, as its name indicates, be used to synthetically and (relatively) realistically overlay a face taken from one video onto another. While this tool could and would be used for fun and political satire, it and related AI media generation technologies quickly spurred public debate about their potential use for disinformation or to violate sexual privacy.[118] On GitHub, these issues were addressed in an informal license included in the *deepfakes_faceswap* repository's README.md file, with the creators stating that 'Faceswap is not for creating inappropriate content [or] any illicit, unethical, or questionable purposes.'[119]

---

[113]Eduard Kovacs, 'Cybersecurity Community Unhappy With GitHub's Proposed Policy Updates' (*SecurityWeek*, April 2021) <www.securityweek.com/cybersecurity-community-unhappy-githubs-proposed-poli cy-updates/> accessed 16 July 2023.
[114]GitHub, 'GitHub Active Malware or Exploits' (*GitHub Docs*, June 2023) <https://github.com/github/docs/blob/cbda3f24344e19678a412f7e9b%5Ctextbackslash%20lware-or-exploits.md> accessed 16 July 2023.
[115]Tarleton Gillespie, 'Do Not Recommend? Reduction as a Form of Content Moderation' (2022) 8(3) *Social Media + Society* 20563051221117552 <https://doi.org/10.1177/20563051221117552>.
[116]See generally David Gray Widder and others, 'Limits and Possibilities for "Ethical AI" in Open Source: A Study of Deepfakes' (FAccT '22, Association for Computing Machinery, 2022) <https://doi.org/10.1145/3531146.3533779>.
[117]Nicholas Diakopoulos and Deborah Johnson, 'Anticipating and Addressing the Ethical Implications of Deepfakes in the Context of Elections' (2021) 23(7) *New Media & Society* 2072 <https://doi.org/10.1177/1461444820925811>.

Nevertheless, the individuals collaborating on *deepfakes_faceswap* clearly understood that it would be used in problematic ways, even creating anonymous alternative accounts to contribute to the tool[120] – an uncommon occurrence on GitHub, where public collaborations are premised on the reputation of the coder and contributions to such a popular repository ('favourited' by more than 45k users as of fall 2023) would normally provide status in the GitHub community.[121] In 2018, GitHub appeared to have gated this repository slightly, making it available for download to logged-in users only, although this restriction seems to have been removed at some point in 2019.[122]

In 2019, *DeepNude*, a deepfake generation system that 'swaps clothes for naked breasts and a vulva, and only works on images of women' was shared on GitHub.[123] After widespread outcry in the tech community, the original creator took down the repository.[124] Nevertheless, several copycat versions were created, and the core Deep-Nude model – a generative adversarial neural network based on UC Berkeley's pix2pix system[125] – was reverse engineered and placed on GitHub (albeit without the previously provided user interface). In contrast to the *deepfakes_faceswap* project, GitHub swiftly removed these models, stating that in response to user flags they 'disabled the project,' which was 'in violation of our acceptable use policy' – with

---

policy staff additionally noting that they 'do not condone using GitHub for posting sexually obscene content and prohibit such conduct in our Terms of Service and Community Guidelines.'[126]

At the time, the relevant part of GitHub's community guidelines stated:

> Don't post content that is pornographic. This does not mean that all nudity, or all code and content related to sexuality, is prohibited. We recognize that sexuality is a part of life and non-pornographic sexual content may be a part of your project, or may be presented for educational or artistic purposes. We do not allow obscene sexual content or content that may involve the exploitation or sexualization of minors.[127]

This policy, perhaps unsurprisingly, seems to struggle with distinguishing between content, content-within-models, and the potential to generate content using a model. As of August 2023, the policy is slightly different, notably stating that GitHub does 'not tolerate content *associated* with sexual exploitation or abuse.'[128] While using this vague term might indicate models that can produce classic 'content' are brought within the policy, it dodges having to distinguish between potential, realised or intended use in this direction. Furthermore, the example of GitHub demonstrates that policy development is only one part of the content moderation pipeline, with enforcement being another crucial dimension. Despite policies that seem to prohibit them, as of fall 2023 multiple Deep-Nude copycat repositories appear to still be online and available on GitHub.[129]

### 3.3. Civitai

Civitai.com quietly went online in the late fall/early winter of 2022. Headquartered in Boise, Idaho, Civitai describes its main offering as a 'community platform dedicated to fine-tuning open-source AI models like Stable Diffusion.'[130] In its inaugural post on the new subreddit r/Civitai, the moderators presented their value proposition:

---

[126]Joseph Cox, 'GitHub Removed Open Source Versions of DeepNude' (*Vice Motherboard*, July 2019) <www.vice.com/en/article/8xzjpk/github-removed-open-source-versions-of-deepnude-app-deepfakes> accessed 16 July 2023.
[127]GitHub, 'GitHub Community Guidelines – GitHub Help' (*GitHub Help (Internet Archive)*, June 2019) <https://web.archive.org/web/20190616104903/>; <https://help.github.com/en/articles/github-community-guidelines> accessed 16 July 2023.
[128]GitHub, 'Sexually Obscene Content Policy' (*GitHub Docs*, June 2023) <https://ghdocs-prod.azurewebsites.net/%5C_next/data/0jDciJzofry75%5C_1LbQ5YT/en/free-pro-team@latest/site-policy/acceptable-use-policies/github-sexually-obscene-content.json?versionId=free-pro-team%5C%40latest%5C&productId=site-policy%5C&restPage=acceptable-use-policies%5C&restPage=github-sexually-obscene-content> accessed 16 July 2023 (emphasis added).
[129]See <https://github.com/topics/deepnude> (archived at <https://perma.cc/AV83-V2P9>).
[130]Crunchbase, 'Civitai – Crunchbase Company Profile & Funding' (*Crunchbase*) <www.crunchbase.com/organization/civitai> accessed 30 October 2023.

> Our users can upload and share custom models that they've trained using their own data, or browse and download models created by other users. These models can then be used with AI art software to generate unique works of art.[131]

The core novel feature provided by Civitai from the onset was an easy-to-use web interface, which not only features various models available for download and easy deployment, but also large thumbnails depicting images generated by each model. The platform has a few features distinguishing it from other model marketplaces: it directs users to a third-party service where they can not only easily prompt a cloud-hosted instance of Stable Diffusion or other open-source image generation models (as through Midjourney's Discord bot, or Stable Diffusion's own ClipDrop web interface) but also share the output content (and tweaked models, if users choose to get more technical) easily with others on the Civitai platform. It also integrates familiar 'social' design elements: interfaces through which logged in users could rate, 'like,' and comment on models and images generated via those models.

As users on Reddit discussed following Civitai's launch, the end-result was a platform far more accessible for image generation tasks than Hugging Face (even though many of the same base models were also available there). One commentator noted that 'you cant [sic] see the [visual output of the] models or anything [on Hugging Face]. You can see instantly at a glance what models are what on Civitai.' Another user noted that 'Civitai offers a better overall experience with its model filtering and easy access to version information and creators' when compared to general-purpose model marketplaces like Hugging Face or Replicate.[132] The end result has been a popular product, with the platform appearing to have tens of thousands of tuned models shared via its platform in August 2023, and an estimated 25-30 million monthly web visits.[133] This success led Civitai to receive a modest but notable 5 million USD funding round in 2023,[134] led by the venture capital firm of Netscape Navigator tycoon and provocateur Marc Andreessen,[135] who had recently published a manifesto decrying 'trust and safety' and 'risk management' as part of a 'self-imposed labyrinth of pain' preventing innovation in the tech sector.[136]

---

[131]See discussion at <www.reddit.com/r/civitai/comments/108gc1k/welcome_to_the_official_civitaicom_subreddit/> (archived at <https://perma.cc/3VCD-KFMV>).
[132]Both comments available at <www.reddit.com/r/StableDiffusion/comments/10rtp17/civitai_alternatives/> (archived at <https://perma.cc/KVX7-WW5U>).
[133]Apparently, according to the same web traffic analysis provider, more than Hugging Face; see Similarweb (n 65).
[134]Pitchbook, 'Civitai Company Profile: Valuation, Funding & Investors' (November 2023) <https://pitchbook.com/profiles/company/530262-28> accessed 30 October 2023.
[135]Emanuel Maiberg, 'Andreessen Horowitz Invests in Civitai, Which Profits from Nonconsensual AI Porn' (*404 Media*, 14 November 2023) <www.404media.co/andreessen-horowitz-invests-in-civitai-key-platform-for-deepfake-porn/> accessed 16 November 2023.
[136]Marc Andreessen, 'The Techno-Optimist Manifesto' (16 October 2023) <https://a16z.com/the-techno-optimist-manifesto/> accessed 16 November 2023.

### 3.3.1. Constructing a content policy

Despite their early success entering the AI art space, the going was not all smooth for Civitai. At the onset, the platform appeared to feature virtually no moderation. Conversations with users on Reddit suggested that the platform was a real bootstrapped operation, run by developers without much (or any) policy counsel. In January 2023, posts made by users on the Civitai sub-Reddit recommended that the company publish a privacy policy and a DMCA copyright takedown request form ('solid idea,' the official Civitai account replied).[137] When urged to interrogate the possible legal liabilities and responsibilities that they had for content created and shared via their services, Civitai staff shrugged, noting that their initial terms of service were 'copied from standard hosting TOS [terms of service]' and 'none of us are lawyers.'[138] Users on the much bigger r/StableDiffusion forum (300k + members) complained about the platform's lack of governance:

> Civitai started as a good idea but it feels like it's been overrun by horny teenage boys. There's also a few questionable models on there lately with underage looking girls. I feel like this site is bad for AI in general and doesn't give a good impression so I want to get away from using it.[139]

Facing this criticism (and perhaps seeking to burnish their reputation as they sought venture funding) Civitai launched a more extensive set of content policies in mid-2023 (Figure 4). The platform now explicitly claimed to not permit a wide range of illegal and/or unsavoury content, including images depicting sexual violence, child abuse imagery (including the cartoon 'shota' subgenre), and hate/discrimination. The policy linked to a few detailed sub-policies, including on models and images intended to depict 'photorealistic minors' and impersonate 'real people.' The new policy on impersonation was promising: it noted that 'Portraying real people in any mature or suggestive context is strictly prohibited,' and that depicting real people (such as celebrities) in anything other than 'conservative high school dress code' would be not permitted.[140] The platform also required users to create an account before viewing potentially NSFW images or models.

To enforce these policies at scale, the content policy live as of Fall 2023 notes that the platform uses a combination of human community moderation and automated classification, noting '[e]very image uploaded is sent for AI content and moderation tagging (utilising Amazon Rekognition), which automatically flags images potentially containing mature content.'[141] For models 'marked as representing' either a real person or a minor, newly generated images using that model will be 'reviewed by a human moderator before being allowed on the site.'[142] The policy appeared to be having some effect, at least insofar as the Civitai forum was inundated with complaints that models and associated images were being removed. In November 2023, Civitai altered its terms of service to cover aspects of model use rather than just uploaded content, specifying that '[a]ll content rules […] apply equally to images generated onsite' – seemingly regardless of whether output content is posted or not.[143] However, this still did not get at the crux of the challenge – the conditions under which the *tools* capable of such on-site generation would be subject to moderation. As one user noted, the new policy efforts just meant that they would need to take more of their image generation offline by downloading models in case they were removed later:

> 'Spent the last hour going through my liked models and downloading all the celebrities. My guess is something spooked them and they're taking down any 'real person' photos showing an inch of skin. There is something called fan art. This is going to be a slippery slope. I have too many other liked models it would take me days to download everything, but I won't be sleeping on downloading anything anymore.'[144]

---

[137]<www.reddit.com/r/StableDiffusion/comments/101j73s/comment/j2ov3hz/> (archived at <https://perma.cc/2SZK-YZX8>).
[138]See <www.reddit.com/r/StableDiffusion/comments/101j73s/comment/j2ouu5k> (archived at <https://perma.cc/8BTF-KE5S>).
[139]<www.reddit.com/r/StableDiffusion/comments/10rtp17/civitai_alternatives/> (archived at <https://perma.cc/KVX7-WW5U>).
[140]Civitai, 'Rules: Real People' (November 2023) <https://perma.cc/3YYD-PGX3> accessed 5 November 2023.

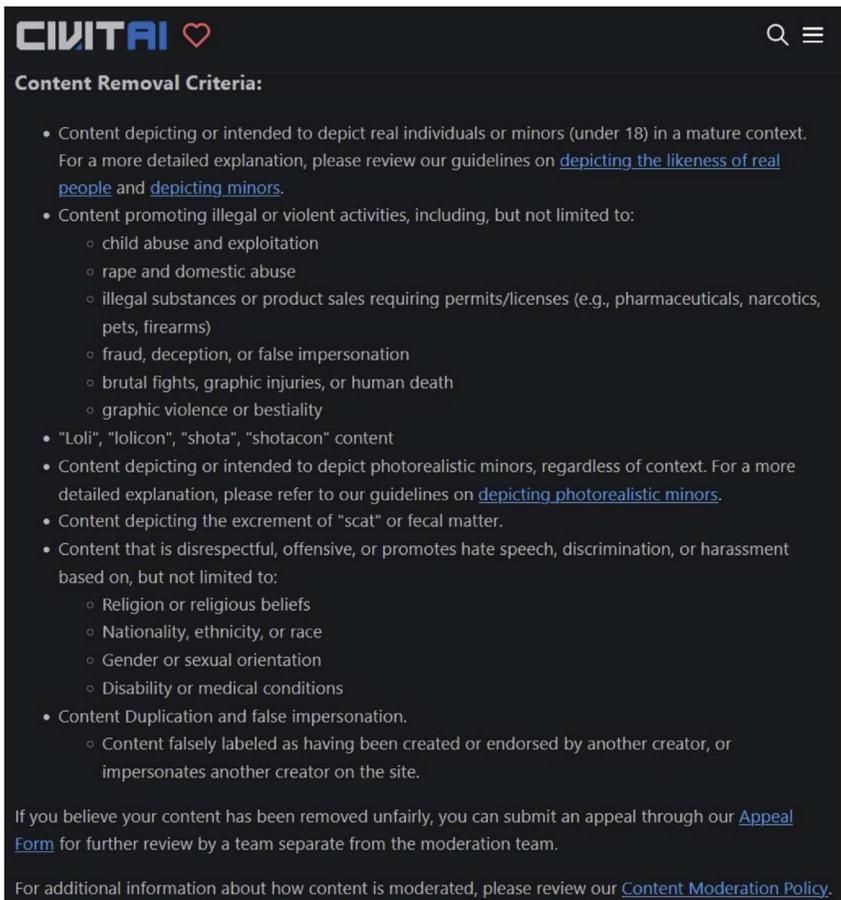

**Figure 4.** Updated Civitai content policy, August 2023.

---

[141]Civitai, 'Rules: Resources' (November 2023) <https://perma.cc/7EGS-RGB5> accessed 5 November 2023.
[142]Civitai, 'Rules: Minors' (November 2023) <https://perma.cc/CST3-CPNY> accessed 5 November 2023.

### 3.3.2. The synthetic porn problem

While a platform like Civitai can certainly be used to produce fan art, cartoon characters, new superheroes, and other creative material, the reality is that a substantial proportion of their traffic is driven by users seeking to create and consume pornography.[145] A 2023 investigation by *404 Media*, an independent tech journalism website, found that Civitai hosted many models explicitly designed to produce pornography, many of which were trained without consent on images scraped from Reddit communities,[146] including those which involve creators posting their own amateur content.[147] As the platform also hosts many models trained on images of celebrities or other real people in an effort to reproduce their likeness, Civitai thus makes it easy to combine these models and create non-consensual pornographic material – broadly understood to constitute a form of image-based sexual abuse – even although this is technically against the platform's content guidelines. While the platform claims to flag some repositories for manual approval before upload, it only seems to do so if the uploader decides to tag the model as 'representing a real person' or 'representing a minor.'[148] Suffice it to say that it seems unlikely that anyone intentionally creating synthetic CSAM will voluntarily submit their material for moderation.

---

[143]Civitai, 'Terms of Service' (November 2023) <https://perma.cc/W9KH-NQ7S> accessed 5 November 2023.
[144]Civitai, 'Rules: Resources' (n 141).
[145]The sheer prevalence of NSFW models on Civitai has led to frequent memory about the platform amongst users of the bigger StableDiffusion subreddit (see for instance <https://perma.cc/5QK7-ZL36>). An informal analysis of the top tags associated with Civitai model pages uploaded by a reddit user suggests that creating images of 'girls' is the most frequent activity for Civitai users; see archive at <https://perma.cc/Z8GL-HDFM>.
[146]Emanuel Maiberg, 'Inside the AI Porn Marketplace Where Everything and Everyone is for Sale' (*404 Media*, August 2023) <www.404media.co/inside-the-ai-porn-marketplace-where-everything-and-everyone-is-for-sale/> accessed 24 August 2023.
[147]Emily van der Nagel and Jordan Frith, 'Anonymity, Pseudonymity, and the Agency of Online Identity: Examining the Social Practices of r/Gonewild' (2015) 20(3) *First Monday*.
[148]Civitai, 'Rules: Real People' (n 140); Civitai, 'Rules: Minors' (n 142).

While some of the examples illustrated by *404 Media* are obvious failures of enforcement – failing to act quickly to remove models tuned to depict nude celebrities like Billie Eilish – the Maiberg[149] investigation reveals how even better resourced actors than Civitai may face issues due to the ability of users to use model training tricks to hide their intent. One example of this involves coded textual inversions seeking to pull out specific features and likenesses from a dataset without explicitly stating that they are doing so. Concerningly, a platform like Civitai centralises, and makes easily accessible from a technical standpoint, the tools for an ex, stalker, abuser, or other motivated person to generate non-consensual images of regular individuals if they have enough photos to tune a model with. It even offers a 'bounty' system where users can create financial incentives for others to create models that can synthesise specific, real people.[150]

### 3.3.3. The bad actor problem

Media reporting suggests that Civitai is at least semi-interested in engaging in some forms of platform governance, removing content that features the most egregious violations of their policies especially when compared to other, even more explicit image-generation model marketplaces.[151] Their founders keep a low profile, however, refusing to speak with the media, not listing any concrete details about their company or themselves on the Civitai platform (which does not have an 'about' or 'contact' page), and seem hesitant to take more intensive actions against their community members – such as removing offending users rather than just models or content generated with those models.

In some areas, such as copyright, the company has taken what can only be described as a highly adversarial stance in relation to creators and other third-parties potentially impacted by their services. For instance, Civitai staff directly emailed an artist, SamDoesArts, who complained about a model impersonating his artistic style, crowing that his complaint spurred a 'Streisand effect' leading to many more users putting together similar models.[152] The email, which was shared publicly by the firm itself on the Civitai sub-Reddit, gleefully informs the artist that Civitai would not only leave these models online, but would actually organise a contest urging others to impersonate 'SamDoesArts,' with voting and prizes for the 'best' impersonations (Figure 5). Civitai staff later confirmed that they 'meant what [they] said' with this 'snarky retort' but just wanted to 'further the conversation.'[153]

---

[149]Emanuel Maiberg, 'Inside the AI Porn Marketplace Where Everything and Everyone is for Sale' (*404 Media*, August 2023) <www.404media.co/inside-the-ai-porn-marketplace-where-everything-and-everyone-is-for-sale/> accessed 24 August 2023.
[150]Emanuel Maiberg, 'Giant AI Platform Introduces 'Bounties' for Deepfakes of Real People' (*404 Media*, 13 November 2023) <www.404media.co/giant-ai-platform-introduces-bounties-for-nonconsensual-images-of-real-people/> accessed 16 November 2023.
[151]Maiberg, 'Inside the AI Porn Marketplace Where Everything and Everyone is for Sale' (n 149).
[152]Jeremy Nuttall, 'Whose Art is This, Really? Inside Canadian Artists' Fight against AI' (*Toronto Star*, February 2023) <www.thestar.com/news/canada/whose-art-is-this-really-inside-canadian-artists-fight-against-ai/article%5C_54b0cb5c-7d67-5663-a46a-650b462da1ad.html> accessed 28 October 2023.

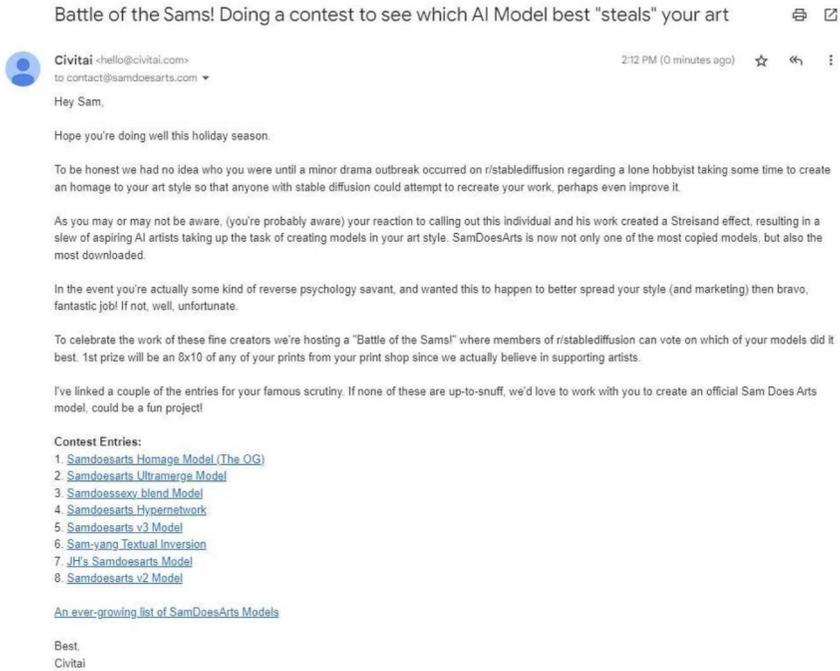

**Figure 5.** The email sent to SamDoesArts from Civitai. Image source: https://www.reddit.com/r/StableDiffusion/comments/zeizyj/battle_of_the_sams_which_samdoesarts_model_does/ posted by user *Slidehussle* (who at least at the time was a Civitai representative and appeared to control the *hello@civitai.com* email account, see https://perma.cc/EMZ3-4CA2.

What will happen if a model marketplace consistently fails to engage in its safety responsibilities, seeking to instead to do only the bare minimum required under laissezfaire American liability frameworks? Or if niche platforms similar to Civitai totally refuse to moderate some forms of harmful content, whether it be illegal (child abuse imagery), and damaging to the livelihoods (copyright) or sexual privacy (non-consensual intimate image abuse) of others? Indeed, Civitai recently launched a new feature, 'Vault,' a form of private storage seemingly designed to resist content moderation. 'Vault,' the firm proudly advertises, 'ensures that even if your favorite resource is deleted from Civitai you retain access.'[154]

---

[153]See discussion at <www.reddit.com/r/StableDiffusion/comments/zeizyj/comment/izka402/?context=3>, archived at <https://perma.cc/G24U-8D5S>.
[154]Civitai, Memberships, Vault, Link, and Generation Updates' (5 April 2024) <https://civitai.com/articles/4 798/memberships-vault-link-and-generation-updates> accessed 11 June 2024.

Regulating intentionally bad actors provides a classic problem for internet regulation, and the policy playbook is likely to involve informal regulation via chokepoints.[155] Some model marketplaces are accessed via other platforms (app stores, community forums like Reddit or Discord) which could potentially be pressured into action by stakeholders in government and/or civil society. But others, like Civitai, are web based, meaning that their takedown would need to be facilitated by actors in the internet infrastructure ecosystem.[156] It is feasible to imagine malicious future model marketplaces being driven off the open web, making their access more difficult (e.g. only via the Tor network) in a manner akin to notorious hate and harassment sites like Kiwifarms.[157] The wide range of platform governance stakeholders could also get creative in efforts to attack bad-actor business models,[158] making it difficult for them to participate in major ad auctions or receive donations and other payments – a possible option as some model marketplaces seek to monetise.[159]

The playbook for this kind of informal regulation of the worst intermediaries online is somewhat established. But for most model marketplaces – those simply seeking to be commercially viable and make a profit – today's policy ambiguities leave much of their action within their own self-regulatory policy discretion. Resourcing will also be an issue: Civitai is a small firm with fewer than 10 reported employees. Many of these platforms have only emerged in 2023, and their trust and safety practices will surely be shaped by both future legal battles and informal forms of pressure from the policymakers that start paying attention.

## 4. Moderating models: emerging governance practices

As powerful generative AI systems become more easily available via large model marketplaces, platform governance scholars will need to grapple with the question of how those models are moderated by the intermediaries that facilitate public access to them. As we have demonstrated, this is a rapidly developing ecosystem with many different actors and new levels of technical complexity. From these short case studies we can identify four emerging practices of note through which leading marketplaces are shaping this area: *friction and access*; *custom licensing*; *bolting-on mitigation features*; and *open policy development*.

---

[155] Natasha Tusikov, 'Defunding Hate: PayPal's Regulation of Hate Groups' (2019) 17(1-2) *Surveillance & Society* 46 <https://doi.org/10.24908/ss.v17i1/2.12908>.

[156] Christoph Busch, 'Regulating the Expanding Content Moderation Universe: A European Perspective on Infrastructure Moderation' (2022) 27 *UCLA Journal of Law and Technology* 32.

[157] Seán Looney, 'Content Moderation through Removal of Service: Content Delivery Networks and Extremist Websites' (2023) *Policy & Internet* <https://doi.org/10.1002/poi3.370> accessed 30 September 2023.

[158] See generally Robert Gorwa, 'Who are the Stakeholders in Platform Governance?' (2022) 24 *Yale Journal of Law & Technology* 493.

[159] In fall 2023, Civitai announced a token system that would allow users to pay for model training and access various premium features, leading users to speculate that it will begin charging for platform access in various ways. See <www.reddit.com/r/StableDiffusion/comments/179hwuh/psa_the_end_of_free_civitai_is_nigh/>, archived at <https://perma.cc/S9SD-8YV4>. See relatedly Tusikov (n 155).

## 4.1. Friction and access

Model marketplaces seem hesitant to act against models that have any potentially beneficial use cases, and generally prefer to take steps other than complete model removal. Moderation interventions have long been conceived as falling on a wide spectrum from 'soft' to 'hard.'[160] 'Hard' measures might include removal of content, user bans, or even reporting to authorities; 'soft' measures might be removal from homepages or limiting discoverability in search and recommender architectures.[161] Somewhere in the middle sit interventions that do not globally remove the content, but make it more difficulty to access or use. Platforms can introduce 'softer' forms of friction by putting content behind interstitials and clickthrough menus, while 'harder' forms of friction might involve geoblocking or 'allow-lists' where users need to register to see specific content or types of content. When it comes to the moderation of machine learning models specifically, a few techniques for introducing this type of 'soft' moderation are already being widely used by leading platforms.

### 4.1.1. Raising barriers for lay users

Model marketplaces often have features that attract lay users to 'try out' models which could themselves be vectors for abuse. For example, Hugging Face enable 'widgets' on model repositories to allow demo querying of models such as text or image generators, as well as a chatGPT-like interface for select large language models such as Meta's LLaMA 2. These can also be disabled by the platform, introducing a small amount of friction into how lay users can interact with a model. While this may limit harm in situations where harm arises from a combination of opportunism and individual queries, this seems a fairly limited response given both the ease of deployment on third party systems as well as many (but not all) AI-related risks relating to issues of scale.

---

[160]James Grimmelmann, 'The Virtues of Moderation' (2015) 17 *Yale Journal of Law and Technology* 42; Robert Gorwa, Reuben Binns, and Christian Katzenbach, 'Algorithmic Content Moderation: Technical and Political Challenges in the Automation of Platform Governance' (2020) 7(1) *Big Data & Society* 2053951719897945 <https://doi.org/10/ggsfrk>.
[161]Gillespie, 'Do Not Recommend?' (n 115).

### 4.1.2. Disabling programmatic functionality

Model marketplaces often integrate APIs as part of their offering (Table 1). We have seen above that GitHub's malware policy centres on disabling access to APIs in order to disrupt ongoing malware attacks while maintaining the abilities of repositories to be downloaded manually (Section 3.2.2). Similarly, intermediaries such as Hugging Face offer one-click deployment functionality with compute providers such as Amazon and Microsoft, and they could disable it (or turn it off by default before models are reviewed and approved) to make model deployment a little more difficult for non-technical users.

### 4.1.3. Login walls and visibility limits

Model marketplaces have added further friction by restricting the ability for users to interact with or download models without an account. GitHub appeared to have taken the very minimal step of making the *deepfake_faceswap* model available for download only to users with a GitHub account in 2018 (Section 3.2.3). This is something that Civitai is also now doing for much of its NSFW content (Section 3.3.2).

This could also be done more broadly by downranking certain models from content discovery mechanisms, such as the 'home pages' or 'popular' pages. Hugging Face is already explicitly discussing the possibility that models may have their visibility limited under certain conditions in their latest content policies,[162] although without default forms of public-facing transparency about these interventions, it is difficult to gauge the extent to which this strategy is a major part of their emerging trust and safety operations.

### 4.1.4. Verification or clickwrap contractual walls

Hugging Face has also extended the login wall concept with a more unique approach permitting model uploaders to require downloaders to enter into an explicit, model-specific contract with the repository maintainer, including providing their contact details (Section 3.1.2). A similar approach seems to have been deployed by the Google-owned Kaggle platform for the release of the Google Gemma multimodal models. Such approaches might aid in moving liability to users via identifiability, which has long been discussed as a potential response to software-generated risks.[163] However, given that marketplaces do not currently verify the contact details of users, this could be interpreted more as psychological 'friction theatre' than a meaningful form of abuse prevention. However, this verification obligation could be externalised via third-party verification providers or through bespoke agreements made with high profile creators. For example, Hugging Face has allowed certain repositories to only be accessed by those on a dynamic, externally-provided allow-list of emails, a strategy they deployed in partnership with Meta to release the LLaMA 2 models.[164]

---

[162]Hugging Face, 'Content Policy' (n 76).
[163]Zittrain (n 58).

There are limits to friction-based model governance strategies. Insofar as download remains possible, models can be integrated into other tools and pipelines, both easy-to-use 'no-code' environments and in ones with higher technical barriers to entry. Login walls are unlikely to be effective without technologies designed to limit account creation per person, or link accounts to real identities, both of which are challenging and potentially problematic.[165] If verification obligations are to be externalised onto smaller developers/creators, it is unlikely that they will have the resources or motivation to undertake this task in lieu of platform staff. The matter is complicated by the ambivalent and dual-use valences of some potentially harmful models. Indeed, as Hugging Face employees noted in a forum post made discussing the GPT-4chan saga, they 'couldn't identify a licensing / gating mechanism that would ensure others use the model exclusively for research purposes.'[166]

### 4.2. Custom licensing as moderation standards

Limiting or shaping the third-party downstream use of models poses a difficult set of governance questions, and Hugging Face in particular has been leaning on licensing as potential solution. As their platform is commonly used to host and deliver fine-tuned or tailored versions of existing models, their platform in effect becomes home to a chain of licensees who are all using a more generic base model for a more specific purpose. This has already created new and difficult dynamics of entanglement between platform policies and license arrangements, one that the platform already has been using not only for safety-related takedowns, but also to justify politically motivated intervention due to jawboning and stakeholder pressure.

The logic motivating this approach is understandable. On one hand, licenses interplay neatly with normal intellectual property based takedown regimes implemented by platforms when following regimes such as the DMCA. When platform companies are made aware of breaches of intellectual property – such as users' posting of content where they do not hold the copyright, if the users cannot muster a license or statutory exemption that permits them to disseminate it – they are generally obliged internationally to take down the content swiftly, or risk becoming liable for the infringement.[167] Yet efforts to verify a breach of a license relating to *use* as is the case in AI model licensing regimes, pose a particular puzzle that challenge the classic notions of licensing as related to content distribution.

---

[164]Philipp Schmid and others, 'Llama 2 Is Here - Get It on Hugging Face' (18 July 2023) <https://huggingface.co/blog/llama2> accessed 5 November 2023.
[165]Bernie Hogan, 'Pseudonyms and the Rise of the Real-Name Web' in John Hartley, Jean Burgess, and Axel Bruns (eds), *A Companion to New Media Dynamics* (Wiley-Blackwell, 2013).
[166]Lewis Tunstall, 'Conditions for Availability (Thread on the Community Page of GPT-4chan)' (June 2022) <https://huggingface.co/ykilcher/gpt-4chan/discussions/4#62aa2c95eab21ac91bb98e19> accessed 5 November 2023.

By analogy, under the EU intermediary liability regime – first set out in the eCommerce Directive[168] and now restated in the Digital Services Act[169] – a platform must have 'actual knowledge or awareness' of infringing content,[170] as well as fail to swiftly take down the content, before becoming liable themselves.[171] This knowledge must be specific and not general, can be obtained via platform's internal content detection systems or through a third party notice (e.g. from a copyright holder), as long as it precise enough to allow a 'diligent economic operator' to reasonably identify, assess and, where appropriate, act against allegedly illegal content.[172] Such a test is designed to ensure that platforms do not have to carry out a 'detailed legal or factual examination' on every instance of uploaded content,[173] such that platforms are only obliged to remove where such illegality is apparent or manifest, avoiding them becoming 'judge[s] of online legality.'[174]

While this may sound promising (as model marketplaces surely wish to also avoid becoming judges of online legality, and instead would prefer to simply interpret licenses when their violation is apparent or manifest) the conditions in licenses such as OpenRAIL which Hugging Face promote are even more laden with uncertainty. Recall that OpenRAIL seeks to prevent users from deploying a model to infer certain categories of sensitive data about an individual, predict health characteristics for the purposes of insurance pricing, attempt to predict criminality, or synthesise undeclared realistic representations of people or events.[175] It is not immediately clear how a platform is supposed to develop a 'bright line' rule that decides if model outputs, such as those derived from base image generation models licensed under OpenRAIL which are then tuned in order to generate pornography, are 'realistic' or not if they depict someone nude without their consent. This conundrum echoes the classic platform governance puzzle of artistic nudity, perhaps best exemplified by Facebook's rule permitting 'handmade' sexual art.[176]

---

[167]See generally Giancarlo Frosio (ed), *Oxford Handbook of Online Intermediary Liability* (Oxford University Press, 1st edn, 2020) <https://doi.org/10.1093/oxfordhb/9780198837138.001.0001>.
[168]Directive 2000/31/EC of the European Parliament and of the Council of 8 June 2000 on certain legal aspects of information society services, in particular electronic commerce, in the Internal Market ('Directive on electronic commerce') [2000] OJ L178/1 (e-Commerce Directive).
[169]DSA.
[170]DSA, recital 22.
[171]This contrasts with the US regime under Communications Act (1934) §230, which is a significant global anomaly insofar as it exempts intermediaries from liability from most content types in effectively all circumstances.
[172]*L'Oréal and eBay* (n 72).
[173]Joined Cases C-682/18 and C-683/18 *Youtube and Cyando* ECLI:EU:C:2021:503, para 35.
[174]Joined Cases C-682/18 and C-683/18 *Youtube and Cyando* (Opinion of Advocate General Saugmandsgaard Øe) ECLI:EU:C:2020:58, para 187.
[175]'Responsible Artificial Intelligence Source Code License' (n 86).

Because of their scope and operation, OpenRAIL and similar licenses also do little to help platforms navigate the questions of what to do with the hosting of dual-use technologies unless derivative versions of them breaching the license are hosted on the platform too. Even where this is the case, it assumes that the risks come from the fine-tuning rather than the original, base models themselves. For example, the *deepfakes-faceswap* model was issued with provider instructions that it was not to be used for abusive purposes (Section 3.2.3), but no obligation sits with the actor to enforce this condition even if they format it as a license. Licenses, even if obliged by marketplaces to accompany certain categories of model, do not do the work of regulating the upload of base models in the first place. Even assuming certain categories of models are obliged to be accompanied by behavioural use licenses, platforms will struggle to find means to oblige the license issuers to enforce those licenses, leaving their enforcement up to the platform itself. In those situations, licenses are just more expensive, less consistent, and more narrowly applicable forms of content moderation.

If model marketplaces are to accept the upload of models with strange, custom licenses, they will have to regularly come to views on whether such licenses have been breached at various points of an AI system's lifecycle. For example, the license that Stable Diffusion is licensed under (the CreativeML Open RAIL-M license) is already vastly different in its prohibitions from the newer OpenRAIL licenses, which in turn are very different from the informal and loosely written licenses such as the cited license of the SoftVC VITS Singing Voice Conversion framework that justified the Xi Jinping model take-down (Section 3.1.2). Overall, we find ourselves in a space with a number of complex and novel moderation dynamics when compared to more traditional user-generated content domains – with multiple actors, overlapping licensing regimes, and potentially hard to unravel 'content supply chains' that all have prospective governance implications.

---

[176]See Robert Gorwa, 'Facebook's Moderation Task is Impossibly Large. It Needs Help' (*Wired*, 2018) <www.wired.co.uk/article/facebook-moderation-art-nudity-origin-of-the-world> accessed 12 November 2023. That said, the open-ended ambiguity of a license like OpenRAIL could be advantageous for individuals seeking to, for example, prevent the easy dissemination and deployment of image generation models tuned to produce non-consensual synthetic pornography with their likeness. With the right reporting channels and/or connections to a model marketplace, the user in question could credibly make an argument that the model outputs are unrealistic and thus should be removed under OpenRAIL. Unfortunately, the same ambiguity could probably exploited by commercially or politically motivated actors putting pressure on platforms remove legitimate models (like those that conduct political satire) as well.

### 4.3. Bolting-on mitigation features

In one of the first papers to explicitly examine the intersection of content moderation and generative AI, Google researchers argued that technical interventions seeking to govern model outputs can be grouped into two types of 'filtering,' which occur on the *input* or *output* side.[177] To take and build upon their categorisation, these interventions can be deployed at different stages of the model design and deployment cycle. Deep input-side interventions fundamentally change how the model operates, by changing parameters or by using curated training data. For instance, one approach could involve bringing in a safety goal at the ground-level of foundation model training, perhaps by carefully hand-labelling input data to pursue a specific design goal – or, because that is time-consuming, expensive, and increasingly at odds with the large-scale approaches that have come to be the machine learning community's state of the art, deploying a third-party tool (such as Google Jigsaw's 'Perspective API' toxic-speech detection system) to remove undesirable content from pre-existing training datasets.[178] Another approach explored by researchers has involved bringing in smaller, curated datasets to tune a general-purpose generative model, seeking to make it statistically unlikely that the model reproduces certain types of undesirable outputs.[179] Emerging work seeks to make such designed-in limitations difficult to undo (through e.g. fine-tuning) where the model weights are known, although as it stands such technologies require a clear and defined set of potential harms at training time, which seems like a daunting challenge.[180] Input-side interventions can also be 'shallow,' in that that they do not meaningfully change how the model operates or how it devises statistical calculations, but instead seek to provide guardrails around how users can interact with it after the fact. One example might involve simple-block lists of keywords that seek to prevent users from including slurs or other problematic words to prompt a ChatGPT-like text-generation model.

A model developer (or platform hosting a model) could also seek to govern model use through output-side interventions. These types of approaches can be understood as those that do not directly seek to shape the calculations made by a model, but instead, allow the model to function normally and produce the requested output – taking this output and processing it in a way that could be used to inform a future governance decision. The current technical tools available here are in most cases exactly the same systems for automated content moderation deployed by most user-generated content platforms: predictive classifiers and hash-based matching systems.[181] Another much discussed type of output-side processing involves watermarking techniques intended to make synthetic content easier to detect.[182]

---

[177]Hao and others (n 12).
[178]Johannes Welbl and others, 'Challenges in Detoxifying Language Models' in Marie-Francine Moens and others (eds), *Findings of the Association for Computational Linguistics: EMNLP 2021* (Association for Computational Linguistics, November 2021) <https://doi.org/10.18653/v1/2021.findings-emnlp.210>.
[179]Such as those which contain personally identifying information or copyrighted material, see Eldan and Russinovich (n 43).
[180]Henderson and others (n 14).

Such moderation systems could be theoretically built as a component of models when delivered as APIs, or as bundled software. For instance, Hugging Face is now hosting optional modules allowing deployers to automatically integrate provenance data meeting the recently developed 'Coalition for Content Provenance and Authenticity' (C2PA) standard into their workflow.[183] As it stands, these safety features are currently easy to remove if individuals have access to the original model source rather than query a black box through others' infrastructure, as they are typically involve code in the inference pipeline distinct from the model itself.[184] In a different type of intervention, Civitai's content guidelines state that they scan the outputs of models hosted on their platform for potential policy violations with Amazon's Rekognition computer vision API.[185] But these filters are typically going to be forced to handle new content rather than having the luxury of hash-matching old content – a much more difficult, and potentially error-prone undertaking.

In the context of content moderation, much automated classification is simply not very good, particularly when faced with nuance and context, and is chock-full of misclassifications at scale.[186] Already, the public-facing Civitai forums are full of users complaining about decisions that they assume could only have been made in an automated fashion, and that output filtering often over-removes legitimate models while leaving many seriously problematic ones (e.g. those depicting children) online. While output-side moderation might help limit misuse by lay users, for example by targeting bundled services like apps or automatic deployment that makes such platforms especially easy-to-use (e.g. Hugging Face's 'Inference Endpoints,' Section 2.1), the efficacy of these interventions is limited by the fact that deployers can generally simply take the published model off-platform.

---

[181] Gorwa, Binns, and Katzenbach (n 160).
[182] Franziska Boenisch, 'A Systematic Review on Model Watermarking for Neural Networks' (2021) 4 *Frontiers in Big Data*; Preetam Amrit and Amit Kumar Singh, 'Survey on Watermarking Methods in the Artificial Intelligence Domain and Beyond' (2022) 188 *Computer Communications* 52 <https://doi.org/10.1016/j.comcom.2022.02.023>.
[183] Alicia Hurst, 'Making AI-Generated Content Easier to Identify' (*Hugging Face*, October 2023) <https://huggingface.co/blog/alicia-truepic/identify-ai-generated-content> accessed 7 October 2023.
[184] Henderson and others (n 14).
[185] Civitai, 'Terms of Service' (n 143).
[186] Gorwa, Binns, and Katzenbach (n 160); Carey Shenkman, Dhanaraj Thakur, and Emma Llansó, 'Do You See What I See? Capabilities and Limits of Automated Multimedia Content Analysis' (Centre for Democracy and Technology, Washington, DC, 2021) <https://cdt.org/insights/do-you-see-what-i-see-capabilities-and-limits-of-automated-multimedia-content-analysis/> accessed 25 May 2021; Inioluwa Deborah Raji and others, 'The Fallacy of AI Functionality' in *2022 ACM Conference on Fairness, Accountability, and Transparency* (2022) <https://doi.org/10.1145/3531146.3533158>.

### 4.4. Open policy development

Emerging content moderation practices on model repositories have often blended together elements of what scholars have conceptualised as two different ideal types of content moderation: commercial and community-oriented.[187] For example, the GitHub malware policy was changed in a process that involved community review of a draft (Section 3.2.2). The Civitai policies appear to be in part a chaotic result of disparate suggestions on Reddit (Section 3.3.1).

Individual decisions are also often elaborated upon in public in a way that is unusual in the traditional social media/user-generated content governance space. GitHub has extensively blogged on and discussed its policy around DMCA anti-circumvention tools in relation to specific decisions (see Section 3.2.1). Hugging Face staffers are also making some of their governance decisions in public on repository talk pages, where other researchers and community members can weigh in with their opinions. In the case of GPT-4chan (Section 1), the model creator had an opportunity to explain his rationale and defend his project from criticism. Many senior executives at Hugging Face posted extensively on the GPT-4chan repository talk forum in the lead up to, and following, the moderation decision made by the company.

While community feedback on policies can scale to even large platforms, open (or at least semi-open), high-level discussions of all controversial individual decisions will likely not. The Xi Jinping takedown requests discussed above (Section 3.1.2) do not appear to have been publicly discussed by Hugging Face staff in this way: unlike the example of GPT-4chan, where download features were removed but the model page itself still is alive, the repository and talk pages associated with the satirical singing models have been completely removed.

---

[187]The classic distinction is that community moderation is done by people (usually volunteers) that are themselves active participants in a community – think of Reddit moderators, Mastodon server administrators, or sysadmins on classic internet bulletin boards. See Casey Fiesler and others, 'Reddit Rules! Characterizing an Ecosystem of Governance' in *Twelfth International AAAI Conference on Web and Social Media* (ICWSM '18', 2018) <https://doi.org/10.1609/icwsm.v12i1.15033>; J Nathan Matias, 'The Civic Labor of Volunteer Moderators Online' (2019) 5(2) *Social Media + Society* 2056305119836778; Joseph Seering and others, 'Moderator Engagement and Community Development in the Age of Algorithms' (2019) 21(7) *New Media & Society* 1417 <https://doi.org/10.1177/1461444818821316>. Commercial moderation is instead done by outsiders not part of an online community, paid employees that may either be part of platform 'integrity' or 'trust and safety' staff or outside contractors hired to review content. See Caplan (n 13); Roberts (n 7); Klonick (n 7); Tarleton Gillespie, *Custodians of the Internet: Platforms, Content Moderation, and the Hidden Decisions That Shape Social Media* (Yale University Press, 2018).

A more open governance culture potentially offers an avenue for users to raise broader concerns, including on more fundamental matters, like the question of how powerful base models should be gated (and at what point should problematic downstream uses of a dual-use model have an impact on its accessibility via a model marketplace). How this will scale into tricky, legally-mandated content moderation requirements is much less clear.

## 5. Conclusions: catalysing analytic capacity

Summer has come again for the AI industry and a wave of new and improved 'open' generative text, image, video, and audio generation models have led to a corresponding boom in platforms that are hosting and facilitating access to them. In this paper, we have sought to shed some light on hosting intermediaries in the AI development ecosystem, demonstrating how a new set of 'model marketplaces' are finding themselves in a role where they can act as important rule-makers, adjudicators, and enforcers shaping how leading ML models are deployed by the public.

The story we have described has clear parallels to the eventual emergence of commercial content moderation on social networks during the Web 2.0 era of user-generated and uploaded content. Platform firms improvised as they grew, with an *ad hoc*, organic form of policy development emerging in response to scandal, public pressure, and stakeholder outcry.[188] Given that almost all major online intermediaries operating today need to create 'trust and safety' bureaucracies to handle problematic forms of user-behaviour, and as they become more publicly prominent and important, face demands from regulators, advertisers, and the public to deal with that behaviour,[189] it is unsurprising that emerging AI platforms are facing similar issues – after all, as Evelyn Douek has quipped, 'everything online eventually becomes a content moderation issue.'[190]

It is similarly unsurprising that we are beginning to see a backlash to even these nascent and incomplete forms of platform governance that Hugging Face and other players have been developing in 2023. Take the example of the newly launched model marketplace *Shoggoth*, built on censorship-resistant peer-to-peer technologies and designed explicitly to prevent the moderation and regulatory-access-point potential of the platforms we describe in this paper.[191] These new, explicitly cyberlibertarian platforms echo both the overarching vision and underlying technical architecture enacted by peer-to-peer platform alternatives such as *Nostr* (for microblogging), *Popcorn Time* (for streaming) or *BitTorrent* and *IPFS* (for general file hosting). Indeed, Meta's LLaMA model, initially gated to researchers, was quickly placed onto BitTorrent by those wishing to circumvent the firm's access controls.[192]

---

[188]Rebecca MacKinnon, *Consent of the Networked: The World-Wide Struggle for Internet Freedom* (Basic Books, 2012); Ben Wagner, 'Governing Internet Expression: How Public and Private Regulation Shape Expression Governance' (2013) 10(4) *Journal of Information Technology & Politics* 389; Klonick (n 7); Nicolas Suzor, *Lawless: The Secret Rules That Govern Our Digital Lives* (Cambridge University Press, 2019).
[189]Gorwa, *The Politics of Platform Regulation: How Governments Shape Online Content Moderation* (n 60).
[190]Evelyn Douek and Johanna Weaver, 'Everything is Content Moderation' (*TechMirror*, August 2023).

Our case studies illustrate the challenges that these emerging platforms are sure to face in the coming years, as well as the broad types of interventions and practices that they have been deploying in response. That said, there are other relatively simply interventions that model marketplaces could be doing, for example by seeking to proactively detect problematic repositories 'as content,' scanning for certain keywords in model cards and model descriptions, perhaps building classifiers that flagged certain models for further review, and hiring 'red team' safety staff actively searching out problematic models (and/or content generated by those models, depending on the marketplace design) for removal.

These are labour intensive practices which will face challenges of resourcing and scale as platforms grow. The future moderation needs of model marketplaces are still unclear: while their userbase will likely remain small compared to social media platforms, and the content base may remain commensurate to that, the nature of the content is in some ways much more complex, more high-stakes, and certainly very difficult to outsource decisions about. Despite efforts by these platforms to deploy existing automated content moderation offerings to help deal with their emerging trust and safety needs, there is little evidence to suggest that the serious, holistic analysis of models can easily be automated. Platforms will also be forced out of their comfort zone by regulators seeking to apply existing 'legacy' laws. Unless platforms intend to acquiesce to every request, there will need to be reproducible processes and mechanisms to help them determine where to draw the line.

Our analysis suggests a wide array of challenges ahead, without any easy solutions. That said, we believe that the question of whether platforms have the resources to act as a careful, fair and proportionate regulatory access point – in an area which often requires resource intensive analysis in order to make considered judgments – will be particularly important. Both the public and platforms would benefit if model marketplaces were able to robustly increase *the analytic capacity needed for model moderation*. Realistically, this will need to involve externalising at least some of this capacity.

---

[191]'Shoggoth Systems' (November 2023) <https://shoggoth.systems/> accessed 5 November 2023.
[192]James Vincent, 'Meta's Powerful AI Language Model Has Leaked Online — What Happens Now?' (8 March 2023) <www.theverge.com/2023/3/8/23629362/meta-ai-language-model-llama-leak-online-misuse> accessed 17 November 2023.

This is not to say that platforms should be dodging their regulatory responsibilities where they are appropriate, but simply to acknowledge that the stakes are high and these relatively small platforms are unlikely to be able to comprehensively deal with the magnitude of the challenge they are facing. Model marketplaces cannot – and should not – attempt to analyse all the models, for all the harms, all the time, by themselves. But if they cannot necessarily trust third-parties submitting complaints, then under what conditions should they accept the arguments made by an outside entity?

The externalisation of content moderation duties is relatively well established in the traditional user-generated content space, with common industry and legal mechanisms including 'trusted flaggers,' researcher data access and third-party fact checkers.[193] That said, model marketplaces exhibit some unique features that will require another level of governance design thinking. As it stands, for either takedown requests or platform's internal investigations, it is unclear what evidentiary threshold should lead to action. Legally, European intermediary liability shields are lowered when actual knowledge is obtained that would convince a 'diligent economic operator' (Section 4.2). In the classic social media context, it seems unlikely that this threshold would be met when evidence is presented that would require scientific analysis in order to be verified – but this is could be exactly the nature of evidence about model potential that AI intermediaries will sometimes need to act on. Furthermore, these platforms are new, and relatively speaking, far less wealthy than established 'big tech' players – so should they be expected to have enormous capacity, as courts often assume of large, advertising funded social networks? What does this mean for the future of AI notice-and-takedown?

One useful first step for model marketplaces could thus be to create template *evidence packs for model flagging*. Evidence packs would be useful for understanding the type of information that would indicate a model's potential for certain kinds of harm, or which would document misuse of a model off-platform. Much like other *pro forma* tools such as model cards for model reporting,[194] evidence packs for model flagging would outline the form and extent of information needed, in this case for action.

We are sceptical that a platform like Hugging Face should act as the unsolicited enforcer of arbitrary licenses (Section 3.1.2). At the very least, only a pre-approved list of those with terms enforceable in principle should be recognised. However, even if marketplaces choose not to do this under their own volition or in response to a request that did not originate from a right-sholder, they still will need to handle requests from the issuers of licenses who claim they are not being adhered to. As these licenses are becoming complex and broad, a standardised, comprehensive reporting process could help fill the inevitable knowledge gap.

---

[193]Sebastian Felix Schwemer, 'Trusted Notifiers and the Privatization of Online Enforcement' (2019) 35(6) Com puter Law & Security Review 105339 <https://doi.org/10.1016/j.clsr.2019.105339>; Naomi Appelman and Paddy Leerssen, 'On "Trusted"' Flaggers' (2022) 24(1) Yale Journal of Law and Technology 452; Jef Ausloos, Paddy Leerssen, and Pim ten Thije, 'Operationalizing Research Access in Platform Governance' (AlgorithmWatch, 2020) <https://lirias.kuleuven.be/3065402> accessed 16 November 2023.
[194]Mitchell and others (n 33).

These 'packs' could also theoretically vary across complaint type. For example, if courts indicate that there are conditions under which trained models can themselves breach copyright or data protection law for containing or reproducing certain information, then evidence pack methodologies would ideally indicate the types of auditing methods which would indicate such a breach. In this manner, these reporting systems could dovetail with marketplaces' chosen internal content policies and community standards.

In effect, we are arguing that there is no clear silver bullet when it comes to moderating models. Funding and capacity building will naturally help, but the unique affordances of powerful ML systems create challenges that even the best resourced, well meaning, 'diligent economic operator' will struggle with. While this space is still somewhat nascent, it seems clear that successful and sustainable governance in this space will require the eventual formation of a broader accountability ecosystem, and the creation and funding of broader structures for appeals, counter-notice, and oversight. GitHub's example is illustrative here: as discussed above (Section 3.2.1), the platform openly complied with pressure to takedown what they perceived to be legitimate creative activity while also seeking to build public capacity to contest over-removals. Thoughtful research and policy in this space will also surely be needed.

The ensuing regulatory conversations will not be easy, and the ensuing implications for public safety are high. We hope that the history of difficult software platform governance challenges – surprisingly sparsely discussed in the literature until now – will inform the responses developed by model marketplaces going forward in their interactions with governance stakeholders and the public.

## Acknowledgements

The ideas in this paper have benefited from conversations with Mike Annany, Anna Bacciarelli, Evelyn Douek, Vidushi Marda, Christopher Persaud, Daphne Keller, Andrew Strait and Dave Willner. We further benefited from extensive feedback from four anonymous reviewers as part of a non-archival submission to the ACM Conference on Fairness, Accountability and Transparency (ACM FAccT '24), and feedback at the conference in Rio de Janeiro. Thanks to participants in the USC Annenberg MASTS, Berkman Klein Center's 'Rebooting Social Media,' FGV Direito Rio's REG Talks, University of Hong Kong, Universität Tübingen's AI and Law Summer School, and Maastricht University G-Law seminars where this paper was presented, as well as attendees of the University of Amsterdam's IViR lecture series. Along with FAccT, early versions of (parts of) this paper were presented at the 'DSA and Platform Regulation' and 'Global Digital Cultures' conferences at the University of Amsterdam in 2023 and 2024, and the Centre for Technomoral Futures Annual Lecture, University of Edinburgh. After the pre-print of our paper was published, we discussed findings with representatives of GitHub and Hugging Face, and these conversations were useful – thank you to Meg Mitchell, Felix Reda, Yacine Jernite and Peter Cihon in particular. Thanks additionally to the incredible journalists at 404 Media (particularly Joseph Cox, Sam Cole and Emmanuel Maiberg) whose work there and previously at VICE Motherboard hugely informed this paper.

## Disclosure statement

## Funding

This work was supported by the UK Research and Innovation Trustworthy Autonomous Systems Hub [grant number EP/V00784X/1].

## Notes on contributors

*Robert Gorwa* is a postdoctoral researcher at the Berlin Social Science Center. He studies the politics of private and public technology policymaking, with a special interest in platform governance and emerging socio-technical regulatory arrangements in the digital economy.

*Michael Veale* is Associate Professor and Vice Dean (Education Innovation) at the Faculty of Laws, University College London, and a Fellow at the Institute for Information Law, University of Amsterdam. He focusses on producing timely, interdisciplinary work at the point where law, policy and emerging digital technologies collide.